\documentclass{emulateapj}

\usepackage{amsmath}
\usepackage{amssymb}
\usepackage{graphicx}
\usepackage{epsfig}
\usepackage{color}
\usepackage{natbib}

\def\gtaprx {\lower .1ex\hbox{\rlap{\raise .6ex\hbox{\hskip .3ex
	{\ifmmode{\scriptscriptstyle >}\else
		{$\scriptscriptstyle >$}\fi}}}
	\kern -.4ex{\ifmmode{\scriptscriptstyle \sim}\else
		{$\scriptscriptstyle\sim$}\fi}}}
\def\ltaprx {\lower .1ex\hbox{\rlap{\raise .6ex\hbox{\hskip .3ex
	{\ifmmode{\scriptscriptstyle <}\else
		{$\scriptscriptstyle <$}\fi}}}
	\kern -.4ex{\ifmmode{\scriptscriptstyle \sim}\else
		{$\scriptscriptstyle\sim$}\fi}}}

\newcommand{\cutt}[1]{\textcolor{blue}{}}

\newcommand{\Ms}{$M_{\odot}$}

\newcommand{\Ni}{{\ensuremath{^{56}\mathrm{Ni}}}}

\begin{document}

\title{Illuminating the Primeval Universe with Type IIn Supernovae}

\author{Daniel J. Whalen\altaffilmark{1}, Wesley Even\altaffilmark{2}, C. C. Lovekin\altaffilmark{3}, 
Chris L. Fryer\altaffilmark{4}, Massimo Stiavelli\altaffilmark{5}, P. W. A. Roming\altaffilmark{6,7},   
Jeff Cooke\altaffilmark{8}, T. A. Pritchard\altaffilmark{7}, Daniel E. Holz\altaffilmark{9} and Cynthia 
Knight\altaffilmark{3,10}}

\altaffiltext{1}{McWilliams Fellow, Department of Physics, Carnegie Mellon 
University, Pittsburgh, PA 15213}

\altaffiltext{2}{XTD-6, Los Alamos National Laboratory, Los Alamos, NM 87545}

\altaffiltext{3}{T-2, Los Alamos National Laboratory, Los Alamos, NM 87545}

\altaffiltext{4}{CCS-2, Los Alamos National Laboratory, Los Alamos, NM 87545}

\altaffiltext{5}{Space Telescope Science Institute, 3700 San Martin Drive, Baltimore,
MD 21218}

\altaffiltext{6}{Space Science \& Engineering Division, Southwest Research Institute, 
P.O. Drawer 28510, San Antonio, TX 78228-0510}

\altaffiltext{7}{Department of Astronomy \& Astrophysics, Penn State University, 525 
Davey Lab, University Park, PA 16802}

\altaffiltext{8}{Centre for Astrophysics and Supercomputing, Swinburne University of 
Technology, PO Box 218, H30, Hawthorn, Victoria 3122, Australia}

\altaffiltext{9}{Enrico Fermi Institute, Department of Physics, and Kavli Institute for 
Cosmological Physics, University of Chicago, Chicago, IL 60637, USA}

\altaffiltext{10}{Department of Physics and Astronomy, Brigham Young University,
Provo, UT 84602}

\begin{abstract}

The detection of Pop III supernovae could directly probe the primordial IMF for the first time, unveiling the 
properties of the first galaxies, early chemical enrichment and reionization, and the seeds of supermassive 
black holes.  Growing evidence that some Pop III stars were less massive than 100 \Ms\ may complicate 
prospects for their detection, because even though they would have been more plentiful they would have 
died as core-collapse supernovae, with far less luminosity than pair-instability explosions.  This picture 
greatly improves if the SN shock collides with a dense circumstellar shell ejected during a prior violent LBV
type eruption.  Such collisions can turn even dim SNe into extremely bright ones whose luminosities can 
rival those of pair-instability SNe.  We present simulations of Pop III Type IIn SN light curves and spectra 
performed with the Los Alamos RAGE and SPECTRUM codes. Taking into account Lyman-alpha 
absorption in the early universe and cosmological redshifting, we find that 40 \Ms\ Pop III Type IIn SNe will 
be visible out to $z \sim$ 20 with \textit{JWST} and out to $z \sim$ 7 with \textit{WFIRST}.  Thus, even low
mass Pop III SNe can be used to probe the primeval universe.

\end{abstract}

\keywords{early universe -- galaxies: high-redshift -- stars: early-type -- supernovae: general -- 
radiative transfer -- hydrodynamics -- shocks}

\maketitle

\section{Introduction}

The first stars in the universe are thought to form in 10$^5$ - 10$^6$ \Ms\ cosmological halos at  $z \sim$ 
20 - 30.  Unfortunately, because these stars lie at the edge of the observable universe there are no 
observational constraints on their properties.  The original numerical simulations of primordial, or Pop III, 
star formation suggest that they are very massive, 100 - 500 \Ms\ and that they form in isolation, one per 
halo \citep{bcl99,abn00,abn02, bcl02,nu01,on07,wa07,on08,y08}.  Newer calculations have since found 
that some Pop III stars may form in binaries \citep{turk09} or even small swarms of 20 - 40 \Ms\ stars 
\citep{stacy10,clark11,sm11,get11,get12}.  Simulations of UV breakout from primordial star-forming disks 
suggest that ionizing feedback in some cases may limit the masses of the first stars to 20 - 50 \Ms\ \citep{
hos11,stacy12,hos12} \citep[but see also][]{op01,oi02,op03,tm04,tm08}.  However, all these estimates 
must be regarded to be very preliminary, since no simulation has evolved a newly formed Pop III 
protostellar disk all the way to the end of the life of one of its stars with realistic physics \citep[for recent 
reviews, see][]{glov12,dw12}.  Furthermore, the impact of turbulence \citep{schl13}, magnetic fields due 
to the small scale turbulent dynamo \citep{schob12}, and radiation transport on the evolution and stability
of the disk are not well understood.

There have been attempts to constrain the Pop III IMF by modeling the nucleosynthetic imprint of primordial 
supernovae (SNe) on later generations of stars.  This imprint is now sought in the fossil abundance record, 
the pattern of chemical elements found in ancient, dim metal-poor stars now being surveyed in the Galactic 
halo \citep[e.g.,][]{Cayrel2004,bc05,fet05,Lai2008,caffau12}.  Recent simulations indicate that 15 - 40 \Ms\ 
Pop III core-collapse (CC) SNe may have contributed significantly to early chemical enrichment, further 
corroborating the existence of lower-mass Pop III stars \citep{jet09b}.  Some have taken the absence of the 
distinctive odd-even nucleosynthetic pattern of pair-instability (PI) SNe \citep{hw02} in extremely metal-poor 
stars to imply that there were no very massive Pop III stars.  However, evidence of the odd-even effect has 
now been found in high-redshift damped Lyman alpha absorbers \citep{cooke11} and perhaps in a new 
sample of stars from the \textit{Sloan Digital Sky Survey} \citep[\textit{SDSS},][]{ren12}.  It is also now known 
that Pop III PI SNe could easily have enriched later stars to metallicities above those targeted by surveys to 
date \citep{karl08,jw11,chen11}.  Much remains to be understood about how metals from the first stars are 
taken up into later generations by cosmological flows \citep[e.g.][]{get07,chiaki12,ritt12}.  Nevertheless, it is 
clear that low-mass Pop III stars may have been common in the early universe, with profound consequences 
for the character of primitive galaxies \citep{jgb08,get08,jlj09,get10,jeon11,pmb11,pmb12,wise12}, early 
reionization \citep{wan04,ky05,abs06,awb07,wa08a} and chemical enrichment \citep{mbh03,ss07,bsmith09}, 
and the origins of supermassive black holes \citep{bl03,jb07b,brmvol08,milos09,awa09,lfh09,th09,li11,pm11,
pm12,jlj12a,wf12,agarw12,jet13,pm13}.

The best prospects for determining the masses of Pop III stars in the near term lie in detecting their SNe
\citep{byh03,ky05,wet08a,vas12}.  Even though they are extremely luminous \citep{s02}, individual primordial 
stars are still too dim to be found by the \textit{James Webb Space Telescope} (\textit{JWST}) \citep{jwst06} 
or 30-meter class telescopes \citep[although in principle their H II regions could be detected via strong 
gravitational lensing;][]{rz12}. Pop III SNe can be 100,000 times brighter than their progenitors or the host 
galaxies in which they reside, and their masses can be inferred from their luminosity profiles.  The main 
obstacle to their detection is Lyman absorption by neutral hydrogen prior to the era of reionization, which 
absorbs or scatters most photons from these ancient explosions out of our line of sight.  Pop III SNe must 
emit enough luminosity below the Lyman limit to be observed in the near infrared (NIR) locally.  \citet{
wet12b,wet12d,wet12a} recently found that \textit{JWST} will detect Pop III PI SNe at any epoch and the 
\textit{Wide-Field Infrared Survey Telescope} (\textit{WFIRST}) and \textit{Wide-field Imaging Surveyor for 
High-Redshift} (\textit{WISH}) will find them out to $z \sim$ 15 - 20 in all-sky NIR surveys \citep[see][for 
past studies of PI SNe and their detection]{sc05,kasen11,pan12a,pan12b,hum12,det12}.  Their extreme 
brightnesses make PI SNe ideal probes of the earliest stellar populations \citep[see][on the detection of 
the SN 2007bi, a PI SN candidate in the local universe]{gy09,yn10}.

Pop III CC SNe may be more plentiful at high redshifts but they are more difficult to detect because of 
their lower luminosities.  \citet{wet12c} calculate detection limits of $z \sim$ 10 - 15 for 15 - 40 \Ms\ Pop III 
CC SNe for \textit{JWST} for explosion energies of 1 - 2 $\times$ 10$^{51}$ erg, so they can be found in 
primitive galaxies but not in the first star-forming halos.  Core-collapse explosions also cannot be used to 
differentiate between primordial and Pop II progenitors in protogalaxies because their central engines are 
not very sensitive to metallicity \citep{cl04,wh07}.  Such events can therefore trace star formation rates in 
the first galaxies but are not ideal probes of the primordial IMF.

However, some CC SNe, Type IIn SNe, have recently been discovered with luminosities that rival those 
of far more powerful explosions.  SN 2006tf and SN 2006gy, whose bolometric luminosities exceed 10$
^{44}$ erg s$^{-1}$ and are on par with those computed for Pop III PI SNe \citep[see, e.g.,][]{gy12}, 
have now been observationally and theoretically connected to the collision of SN ejecta with a dense 
circumstellar shell ejected by a violent luminous blue variable (LBV) eruption a few years prior to the 
death of the star \citep{nsmith07a,gy07,gy09a,vmarle10,chev11}.  The shell is thought to be opaque 
because only photons from the collision are observed, not those from shock breakout from the surface 
of the star.  Type IIn SNe likely occur at high redshifts because 15 - 40 \Ms\ Pop III stars that do not 
have much convective mixing over their lifetimes are known to die as compact blue giants \citep{sc05,
jet09b} \citep[see also][on the effect of rotation and magnetic fields on the evolution of Pop III stars]{
Ekstr08,stacy11b,yoon12,stacy13}.  They may be observable at redshifts above those at which normal
luminosity CC SNe can be detected because the shell becomes extremely luminous in UV when the 
ejecta crashes into it \citep{moriya10,tet12,moriya12} \citep[see also][]{tomin11}.  

We have modeled Pop III Type IIn SNe and their light curves and spectra with the Los Alamos RAGE 
and SPECTRUM codes in order to calculate their detection limits in redshift and their NIR signatures.  In 
Section 2 we describe our explosion and shell models and how they are evolved in RAGE.  Blast profiles, 
light curves and spectra are examined in Section 3, and in Section 4 we compare light curves from our 
models with those of Type IIn SN candidates observed in the local universe.  In Section 5 we calculate 
NIR light curves for Pop III Type IIne at high $z$ and determine their detection limits as a function of 
redshift.  In Section 6 we conclude.

\section{Numerical Models}

We take the z40G SN from \citet{wet12c} (hereafter WET12) to be our fiducial explosion model.  It is a 
zero-metallicity, 40 M$_{\odot}$, 2.4 $\times$ 10$^{51}$ erg CC SN.  This progenitor was evolved from 
the beginning of the main sequence up to the point of explosion in the Kepler code \citep{Weaver1978,
Woosley2002}, at which point the SN was artificially triggered and followed until the end of all nuclear 
burning.  Profiles for the blast, which at this point was still deep in the star, were then mapped onto a 2D 
grid in the CASTRO adaptive mesh refinement (AMR) code \citep{Almgren2010} and evolved until just 
before shock breakout to capture mixing inside the star due to Rayleigh-Taylor instabilities.  We begin
this study by spherically averaging the final CASTRO density, energy, velocity and mass fractions and 
mapping them onto a 1D spherical AMR grid in RAGE together with the surrounding star, its wind and 
a variety of dense shells.  

We adopted z40G, a red supergiant progenitor rather than a blue compact giant star like those thought
to be the origin of LBV outbursts, because its larger surface area gives an upper limit to the luminosity 
from shock breakout and hence the number of photons that initially escape through the shell.  The 
dynamics of the collision with the shell is not sensitive to the structure of the star.  Had we instead used 
the u40G explosion a greater fraction of the breakout transient would have gotten through the shell 
because more of it would have been x-rays (shocks breaking out of compact stars are hotter than those 
breaking free from red giants for a given explosion energy).  On the other hand, blue stars have smaller 
surface areas and this compensates for higher shock temperatures in the flux.  None of these issues 
impact detection limits because the breakout pulse is completely absorbed by the neutral IGM at high 
redshift.  

\subsection{RAGE}

RAGE \citep[Radiation Adaptive Grid Eulerian;][]{rage} is a multidimensional adaptive mesh refinement 
radiation hydrodynamics code developed at Los Alamos National Laboratory (LANL).  It couples second
order conservative Godunov hydrodynamics to grey or multigroup flux-limited diffusion (FLD) to model 
strongly radiating flows.  Our RAGE root grid has 200,000 zones with a resolution of 4.0 $\times$ 10$^{
10}$ cm and reflecting and outflow conditions on the inner and outer boundaries, respectively.  Our 
choice of mesh ensures that all important features of the shock, the star, and the shell are resolved by 
at least 10 zones and that 5000 zones are allocated from the center of the grid to the outer edge of the 
shock at setup.  As in WET12, we allow up to five levels of refinement so each feature can be further 
resolved by up to 320 additional zones if necessary, and we again periodically resample the explosion 
onto larger grids to accommodate its expansion and speed up the calculation.  All physics in the WET12 
models are used here:  multi-species advection, grey flux-limited diffusion radiation transport with Los 
Alamos OPLIB atomic opacities\footnote{http://aphysics2/www.t4.lanl.gov/cgi-bin/opacity/tops.pl} \citep{
oplib}, 2-temperature physics, and energy deposition due to radioactive decay of \Ni. 

We evolve the explosion through breakout from the star, collision with and propagation through the shell, 
and expansion into the IGM out to 500 days.  When we post-process dumps from RAGE with SPECTRUM 
to obtain light curves, we sample shock breakout from the star with 50 spectra at evenly spaced times that 
bracket the thermal transient and 200 spectra at logarithmically spaced times out to 500 days.  Computing 
spectra in this manner allows us to model with detailed opacities how the shell attenuates both the initial 
breakout pulse and the flash from its own lower layers when the ejecta collides with it.

\subsection{SPECTRUM}

RAGE profiles are imported into the LANL SPECTRUM code \citep{fet12} in order to calculate spectra
with monochromatic OPLIB atomic opacities that capture detailed emission and absorption line structure.
As explained in \citet{fet12}, densities, velocities, and gas and radiation energy densities from the most 
refined levels in the RAGE AMR hierarchy are first extracted and reordered by radius into separate data 
files. These profiles are then mapped onto a 2D grid in radius and $\mu = cos \, \theta$ in SPECTRUM. 
The spectrum calculation is done with 5200 radial bins and 160 angular bins, the same number as in 
WET12 but with a slightly modified gridding strategy.  The simulation volume is divided into 3 regions: 
inside the $\tau=20$ surface, between the $\tau=20$ surface and the radiation front, and beyond the 
radiation front.  The position of the radiation front is taken to be the outermost cell with a temperature 
above 0.0292 eV, approximately three times the background temperature of 0.01 eV (136 K).  The 
radius of the $\tau=20$ surface is calculated from the outer edge of the grid assuming $\kappa =$ 0.2 
cm$^2$/g, that due to electron scattering in gas at primordial composition, 76\% H and 24\% He by 
mass.  To avoid allocating to many zones to the cold dense shell and under-resolving the shock prior 
to its impact with the shell, the $\tau=20$ surface calculation excludes the density of the shell before 
the collision.  If a zone in the shell has not received additional momentum from the shock then the 
density in that cell is replaced with what the wind density would be at that location.  This allows 
SPECTRUM to adequately resolve the photosphere of the ejecta prior to its collision with the shell and 
then the shell itself when it becomes the dominant source of flux.

\subsection{Shell Structure}

\begin{figure}
\plotone{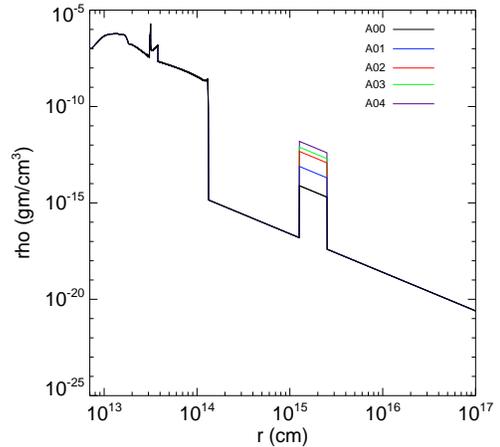}
\caption{Density profiles for the shells considered in our study (models A00 through A04). The SN shock 
is visible at 4.0e13 cm and the surface the 40 \Ms\ star is visible at 1.3e14 cm. \vspace{0.1in}} 
\label{fig:shell}
\end{figure}

\begin{deluxetable}{cccc}  
\tabletypesize{\scriptsize}  
\tablecaption{Pop III Circumstellar Shell Models \label{table:t1}}
\tablehead{
\colhead{model} & \colhead{$M_{sh}$ (\Ms)} & \colhead{$\dot{m}_w$ (\Ms\ / yr)} & \colhead{$v_{w}$ (km s$^{-1}$)}}
\startdata 

A00 & 0.1 & 10$^{-4}$ &  200 \\
A01 &  1   & 10$^{-4}$ &  200 \\
A02 &  6   & 10$^{-4}$ &  200 \\
A03 &  10 & 10$^{-4}$ &  200 \\
A04 &  20 & 10$^{-4}$ &  200

\enddata 
\end{deluxetable}   \vspace{0.2in}

Pop III stars are not normally thought to lose much mass over their lifetimes because there are no line
driven winds in their metal-free atmospheres \citep{Kudritzki00,Vink01,Baraffe01,kk06,Ekstr08}.  
However, these studies do not exclude the possibility of violent pulsational mass loss late in the life of 
massive primordial stars.  For simplicity, we adopt the shell structure used in \citet{vmarle10}, a 
transient, high-mass flux that interrupts the more diffuse wind blown by the star before and after the 
ejection. The two winds have the same constant velocity but different uniform mass loss rates that 
together create a density profile that is a simple superposition of two winds: 
\vspace{0.1in} \[ \rho(r) = \left\{ \begin{array}{ll}
			   \frac{\dot{m}_w}{4 \pi r^2 v_w}  & \mbox{if $r \leq r_{1}$} \, \mbox{and $r \geq r_{2}$} \\
			   						   &										    \\
			   \frac{\dot{m}_{sh}}{4 \pi r^2 v_{w}}  & \mbox{if $r_{1} < r < r_{2}$.}					   
                          \end{array}
                  \right.\vspace{0.1in} \]
Here, ${\dot{m}}_w$ and ${\dot{m}}_{sh}$ are the mass loss rates of the wind and the shell, where 
\begin{equation}
\dot{m}_{sh} \, = \, \frac{M_{sh}}{dt_{sh}}
\end{equation}
and $v_w$ is the wind speed.  The two radii $r_1$ and $r_2$ mark the inner and outer surfaces of the 
shell,
\begin{eqnarray}
r_1 & = & v_w t_{end}  \\
r_2 & = & v_w (t_{end} \,+ \, dt_{sh}),
\end{eqnarray} 
where $t_{end}$ is the time between the end of the ejection and the SN, and $dt_{sh}$ is the duration of the 
ejection.  As in \citet{vmarle10}, $t_{end}$ and $dt_{sh}$ are 2 yr in all our models and we vary the density 
of the shells by adjusting $M_{sh}$.  We take the shells to be primordial, 76\% Hand 26\% He by mass 
fraction. Density profiles for the 5 shells in our study together with the 40 \Ms\ star are shown in Figure 
\ref{fig:shell}, and we summarize the properties of the shells in Table \ref{table:t1}.

The aim of our numerical campaign is to explore the observational signatures and detection thresholds of 
Pop III Type IIn SNe, not perform an exhaustive survey of such explosions.  We adopt this type of shell to 
compare our light curves with those of \citet{vmarle10} and because it is similar in mass and structure to 
the shell inferred to exist around $\eta$ Carinae from observations.  Density profiles for actual LBV 
eruptions are likely more complicated, with fast winds preceding and following much slower outbursts that 
create multiple shocks and rarefaction zones similar to those found in \citet{met12a}.  Radiative cooling 
would also flatten the shell into a colder and denser structure than the ones shown here if dust and metals 
are present.  Type IIn SNe in the local universe will be examined in a forthcoming study.

\begin{figure*}
\begin{center}
\begin{tabular}{cc}
\epsfig{file=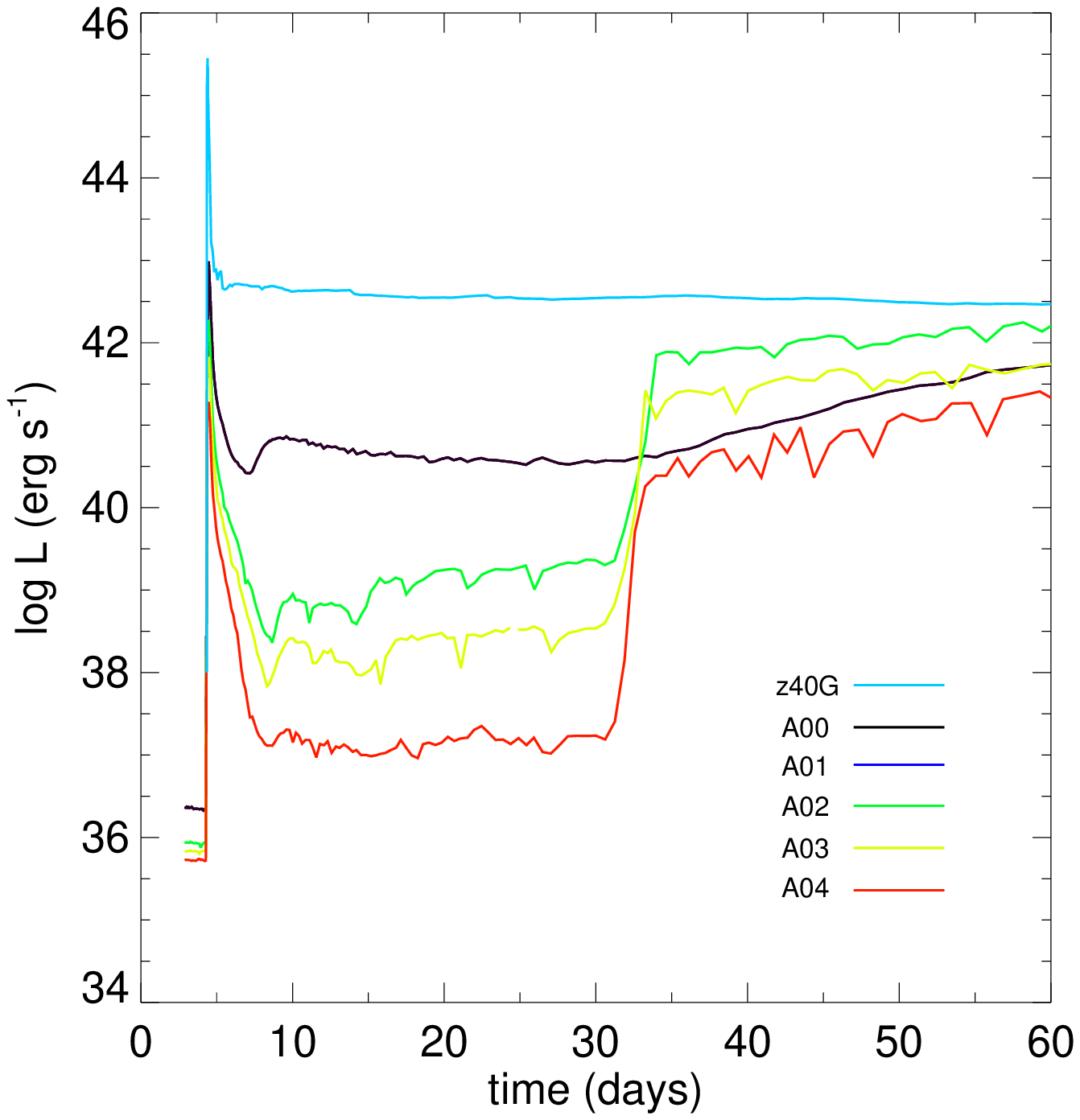,width=0.45\linewidth,clip=} & 
\epsfig{file=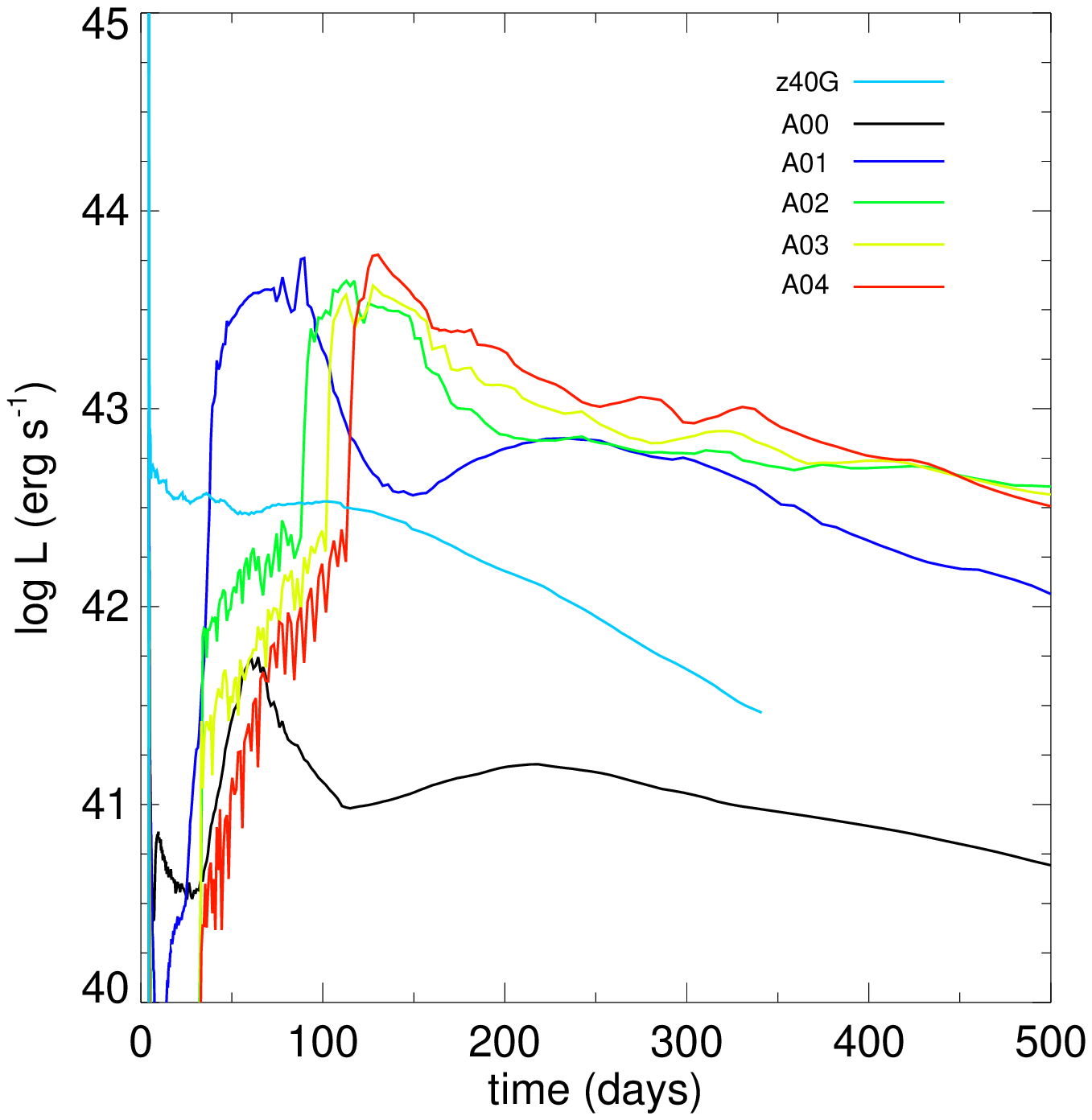,width=0.45\linewidth,clip=} \\
\end{tabular}
\end{center}
\caption{Bolometric luminosities for all 5 Type IIn explosions together with the z40G explosion.  Left panel:  
out to 60 days.  Right panel:  out to 500 days.  Shock breakout from the star itself appears as the luminosity
spike at 4.5 days, the collision of the radiative precursor with the inner surface of the shell is visible as the 
small uptick in luminosity at 9 days, and the collision of the ejecta with the shell causes the jump in luminosity 
at 30 - 40 days.  Breakout of ejecta from the outer surface of the shell is visible as the surge in bolometric
luminosity at 60 - 125 days on the right.}
\label{fig:lc}
\end{figure*}

\subsection{Ionization State of the Shell}

Before launching our runs in RAGE we performed a separate test to determine if UV radiation from the 
star ionizes the shell \citep[see, e.g.,][]{wan04}.  The ionization state of the shell determines its opacity to 
the SN before and after its collision with the shell.  It may also influence how efficiently the kinetic energy 
of the ejecta is transformed into radiation upon impact with the shell, and hence its luminosity. Both issues 
are relevant because the progenitor can be tens of solar masses and therefore a very luminous source of 
ionizing UV photons, and there is little if any dust in the vicinity of the star to attenuate them.

We use the ZEUS-MP code to determine if the star ionizes its shell \citep{wn06,wn08b,wn08a}. ZEUS-MP 
self-consistently solves hydrodynamics, nonequilibrium H and He chemistry and ionizing UV transport to 
propagate cosmological ionization fronts.  We consider a 250 \Ms\ Pop III star in shell A00 from Table 
\ref{table:t1}.  The wind and shell are initialized on a 1D spherical mesh with 200 uniform zones and inner 
and outer boundaries at 1.3 $\times$ 10$^{13}$ cm (the surface of the star) and 3.0 $\times$ 10$^{15}$ 
cm (the outer surface of the shell 2 yr after the end of the ejection). We use multifrequency UV transport, 
with 40 bins uniformly partitioned in energy from 0.255 to 13.6 eV and 80 bins logarithmically spaced from 
13.6 to 90 eV.  The blackbody spectrum of the star is normalized to ionizing photon emission rates, surface 
temperatures, and luminosities from \citet{s02}.  The shell is illuminated for 4 yr, the time from the onset of 
ejection to the SN.  This treatment is approximate, given that the shell is initially closer to the star and 
exposed to higher fluxes just after expulsion.  Because a 250 \Ms\ star is more luminous than a Type IIn 
progenitor, its ionizing flux is the extreme upper limit that could be reasonably applied to the shell.

We find that the radiation front from the star easily ionizes the wind on timescales of a few hours but is 
halted by the shell without ionizing even one zone of it.  Since the star is incapable of ionizing the least
massive shell in our study, we take all the shells in our RAGE models to be neutral.  We note that had 
the shells been ionized they would fully recombine by the time the explosion reaches them.  This is 
evident from the recombination timescales in the gas,
\begin{equation}
t_{rec} = \frac{1}{n_e \alpha(T)},
\end{equation} 
where 
\begin{equation}
\alpha(T) \, = \, 2.59 \times 10^{-13} {T}^{-0.75} \, \mathrm{cm}^{-3} \mathrm{s}^{-1}
\end{equation} 
is the case B recombination rate coefficient for hydrogen, $n_e$ is the electron number density and $T_4$ 
is the temperature in units of 10$^4$ K.   With densities of 10$^9$ cm$^{-3}$ in the A00 shell and ionized 
gas temperatures $T \sim$ 10,000 K, $t_{rec}$ is about an hour.   

\section{Pop III IIn SN Light Curves and Spectra}

\begin{figure}
\plotone{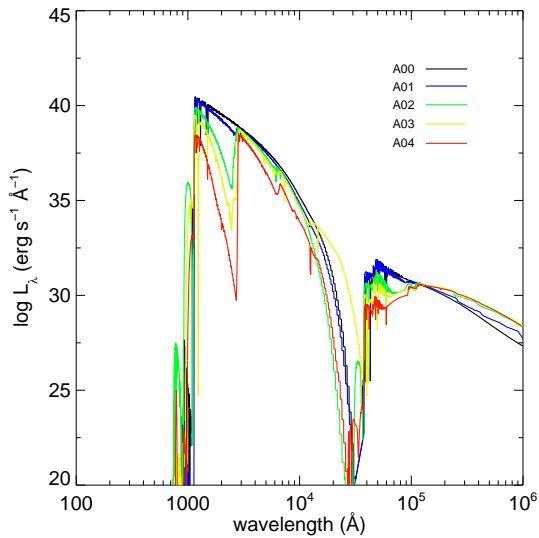}
\caption{Spectra at shock breakout from the surface of the star (4.5 days) for all five Type IIn explosions.
The absorption features at 1000 - 3000 \AA\ due to the shell cause the drop in bolometric luminosity at 4.5 
days.} 
\label{fig:atten}
\end{figure}

We show bolometric light curves for all five explosions in the source frame out to 60 days and 500 days in 
the left and right panels of Figure \ref{fig:lc}, respectively, and spectra for shock breakout from the surface 
of the star in Figure \ref{fig:atten}.  Blast profiles for the A04 explosion are shown in Figures \ref{fig:A04_1} 
- \ref{fig:A04_3}.  The evolution of the SN can be partitioned into three distinct phases:  (1) the collision of 
a radiative precursor with the shell; (2) the collision of the ejecta with the shell and its propagation through 
the shell; and (3), breakout from the outer surface of the shell and expansion into the surrounding medium.

\subsection{The Radiative Precursor}

\begin{figure*}
\begin{center}
\begin{tabular}{cc}
\epsfig{file=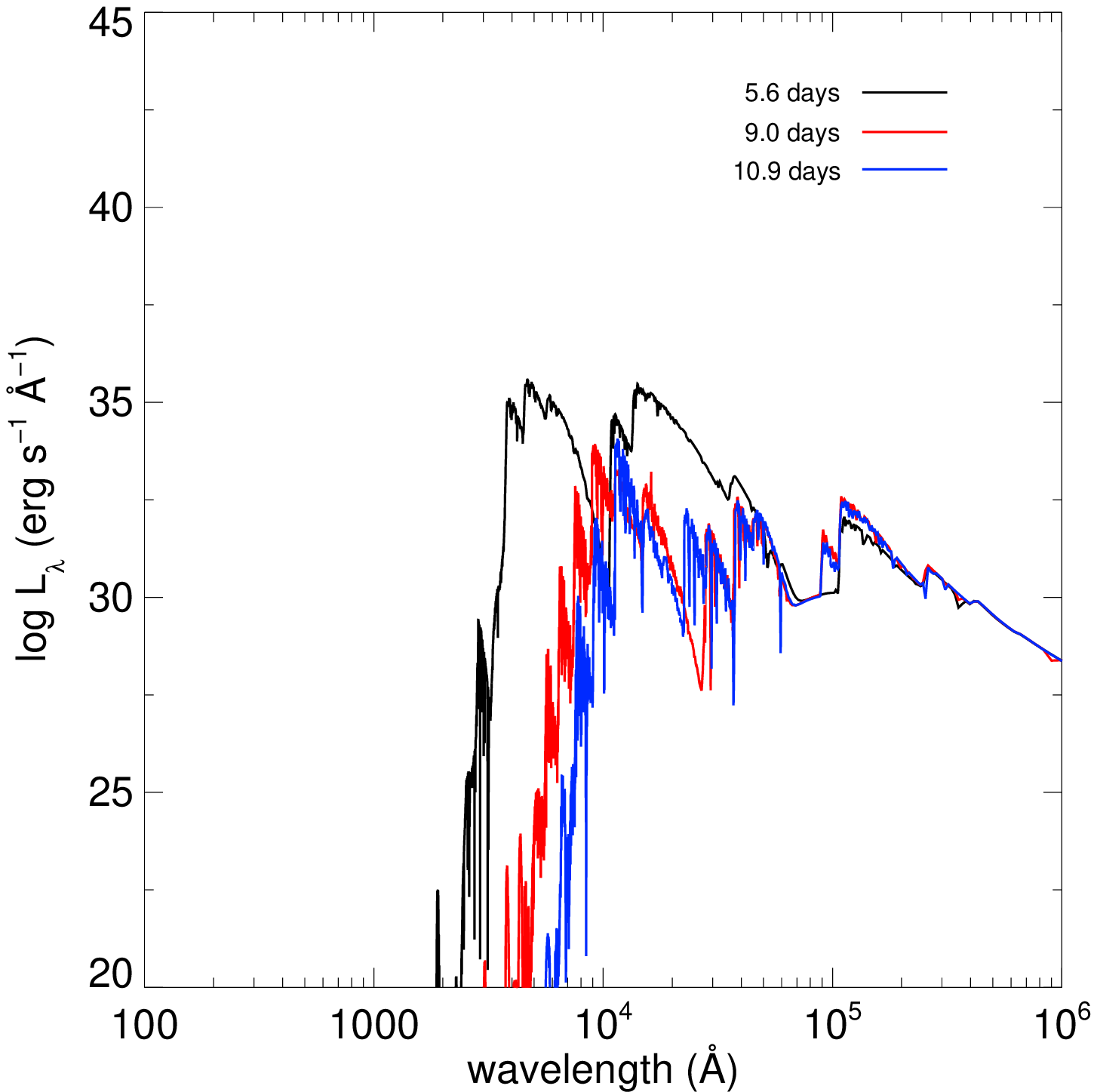,width=0.45\linewidth,clip=} & 
\epsfig{file=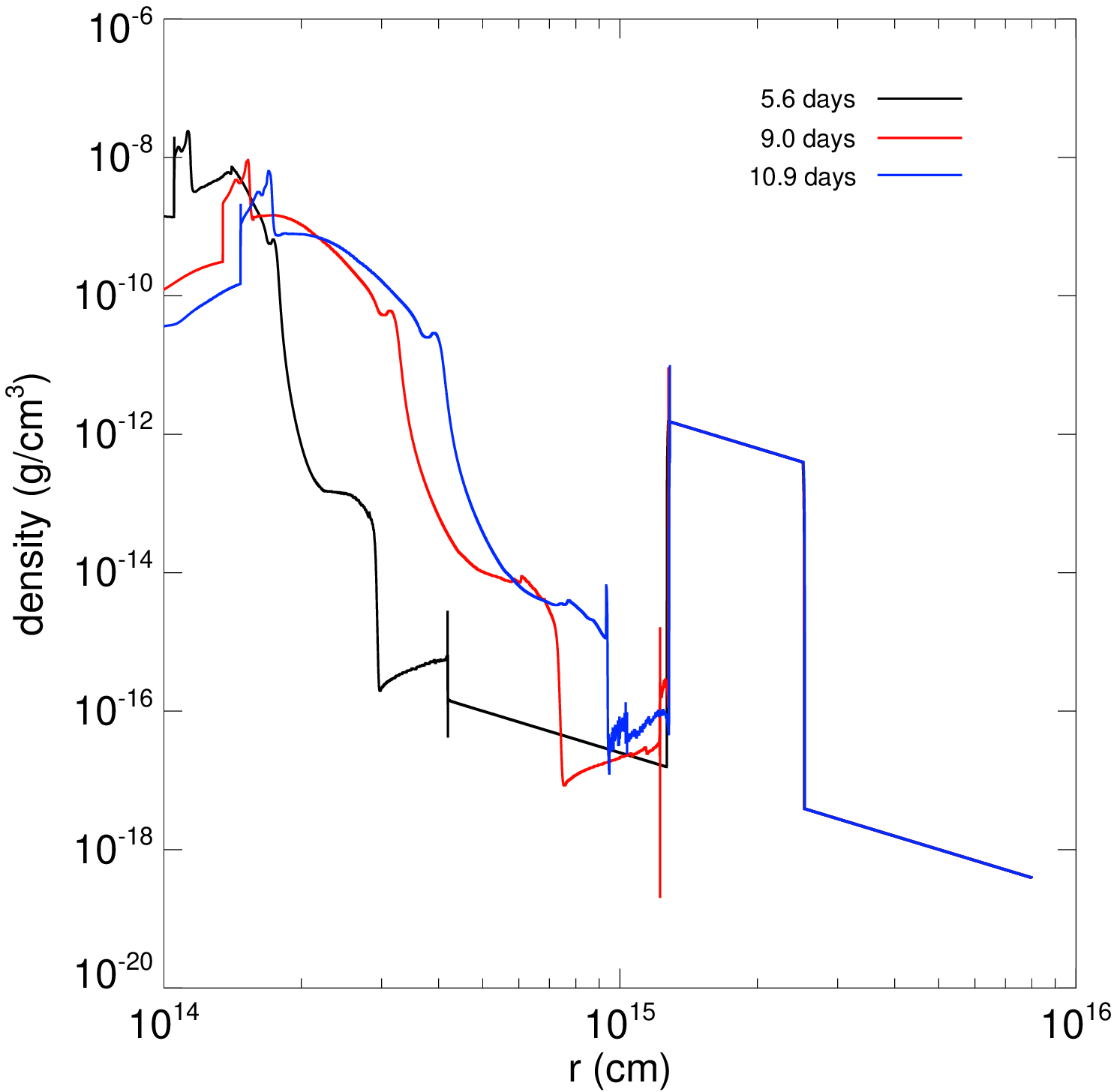,width=0.45\linewidth,clip=} \\
\epsfig{file=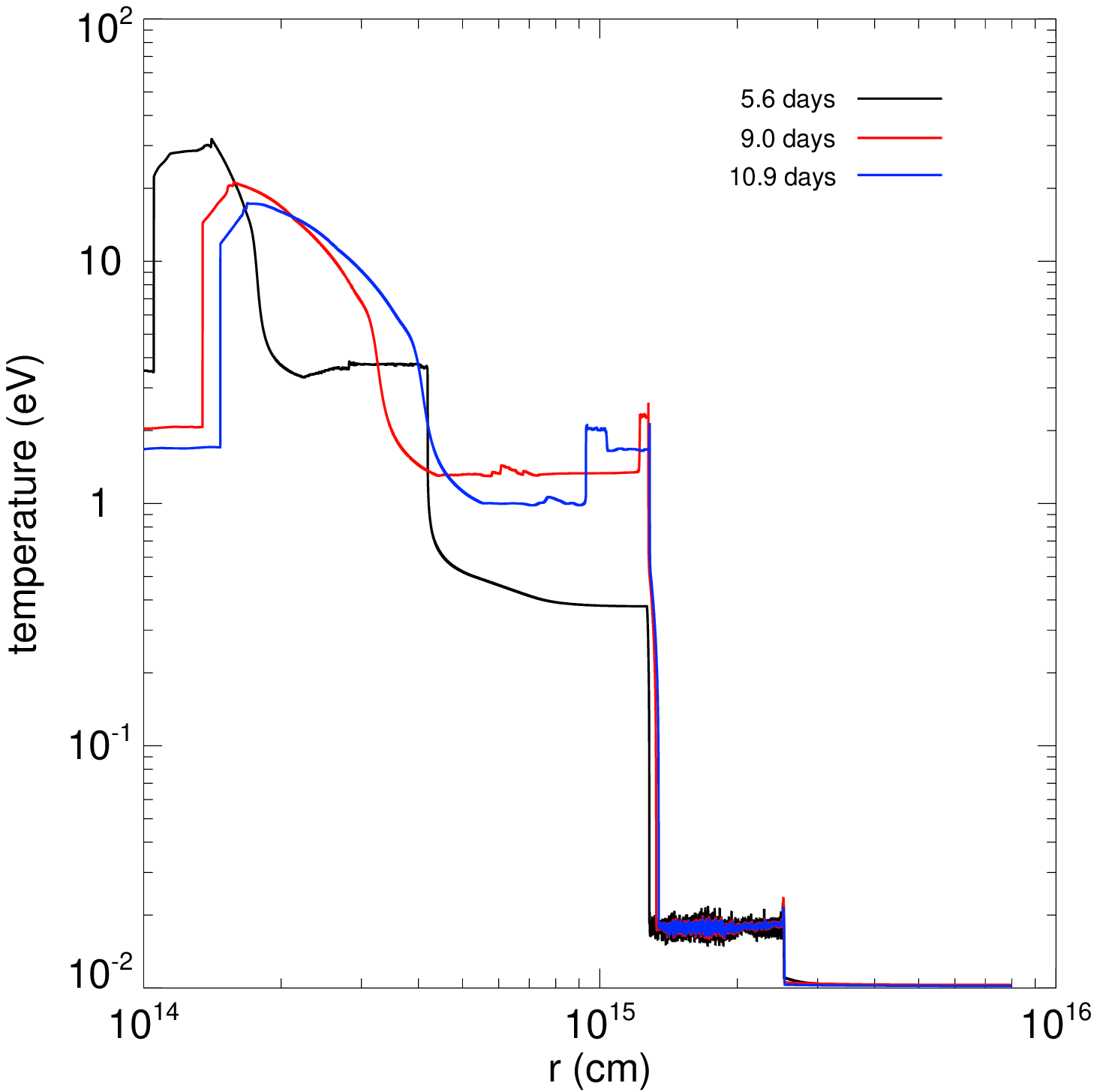,width=0.45\linewidth,clip=} &
\epsfig{file=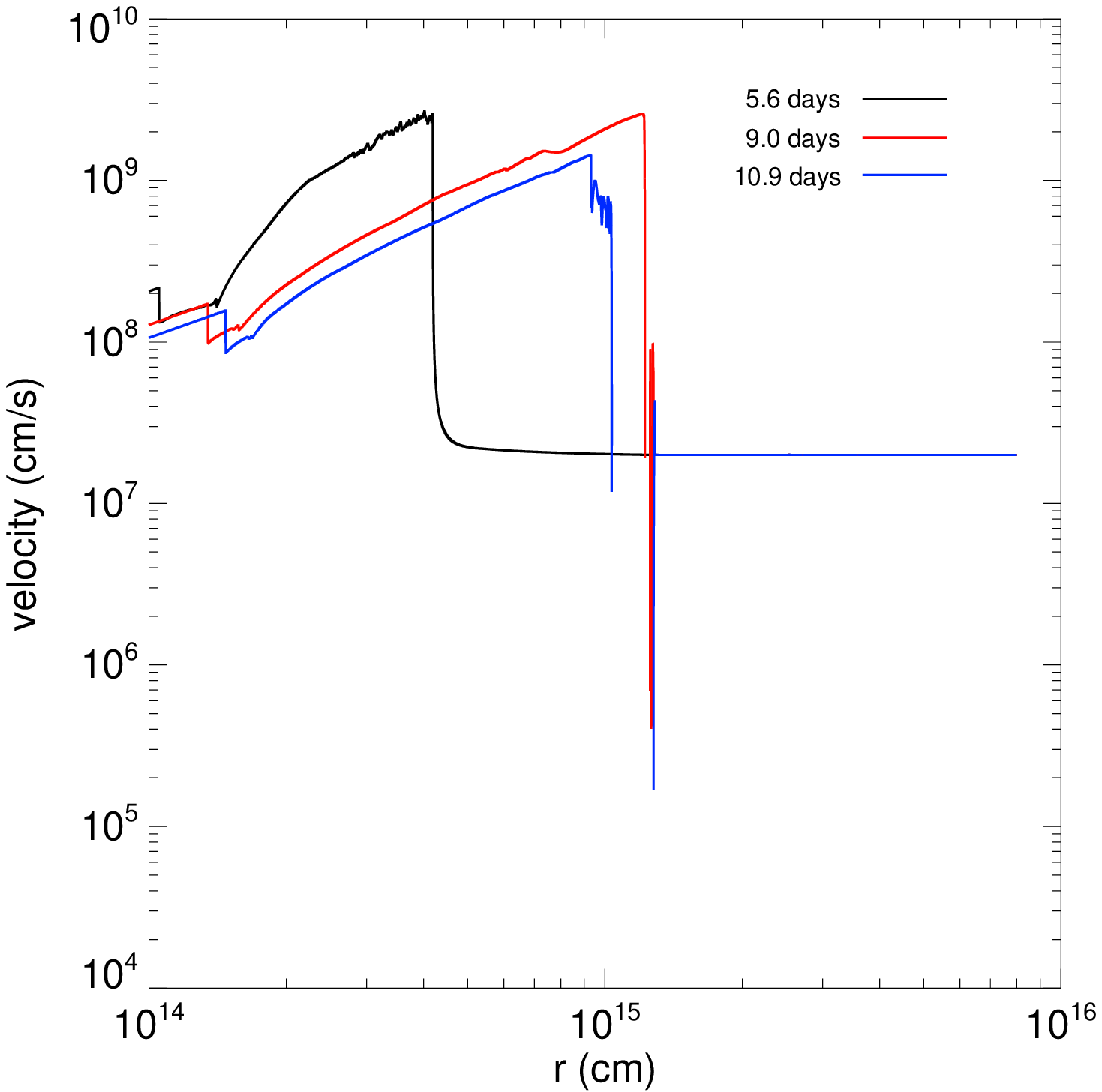,width=0.45\linewidth,clip=}
\end{tabular}
\end{center}
\caption{The collision of the radiative precursor, the outermost layers of the star blown off by radiation 
at shock breakout, with the inner surface of the A04 shell. Upper left panel:  spectra; upper right panel:  
densities; lower left panel:  radiation temperatures; lower right panel:  gas velocities.  As the precursor 
piles up against the inner surface of the shell, a shock (visible as the peaks in radiation temperature at 
9.0e14 cm at 9.0 days and 1.3e15 cm at 10.9 days) reverberates back and forth through the 
accumulated gas.}
\label{fig:A04_1}
\end{figure*}

As shown in the left panel of Figure \ref{fig:lc}, some flux from shock breakout from the surface of the star 
passes through all five shells. The breakout transient, shown at 5 days in the left panel of Figure \ref{fig:lc}, 
is $\sim$ 5 $\times$ 10$^{45}$ erg s$^{-1}$ but is reduced to 10$^{43}$ erg s$^{-1}$ by the A00 shell and 
to 10$^{41}$ erg s$^{-1}$ by the A04 shell.  This attenuated flux is unlikely to be detected because it is 
relatively dim and because the shell becomes much brighter later on. As we show in Figure \ref{fig:atten}, 
most of the attenuation is due to the absorption of 1000 - 3000 \AA\ photons. The bolometric luminosity of 
the shock falls to $\sim$ 10$^{42.5}$ erg s$^{-1}$ after about a day and decays slowly thereafter. 

The photon pulse blows off the outer layers of the star after breaking free of the shock and drives them 
across the gap between the star and the shell.  This radiative precursor, the feature at 4.0 $\times$ 10$
^{14}$ cm at 5.6 days in the density and velocity profiles in Figure \ref{fig:A04_1}, reaches with the inner 
surface of the shell at 9 days, well before the ejecta.  The collision creates a temperature spike at the 
inner surface of the shell that is manifest as the small bump in luminosity just after the breakout transient 
in all five shell light curves.  After the initial collision, the wispy outermost layers of the star continue to pile 
up in a thin layer at the inner surface of the shell, forming a reverse shock that reverberates back and 
forth through the layer.  This can be seen in the temperature profile across the layer: the small spike in 
temperature at 1.3 $\times$ 10$^{15}$ cm, the upper surface of the layer, at 9.0 days rebounds to 9.0 
$\times$ 10$^{14}$ cm, the lower surface, by 10.9 days.  Slightly fewer high-energy photons get through 
the shell at 10.9 days than 9 days because the reverse shock is at the bottom of the thin layer and more 
of its radiation is absorbed.

The reverberation of the shock back and forth in this layer causes the flickering in the light curve from 8 - 
32 days.  The luminosities are relatively steady over these times because the thin layer is trapped at the 
inner surface of the shell.  However, luminosities for less massive shells are higher because they allow 
more radiation to pass.  The ripples in the A02 - A04 light curves are absent from the A00 and A01 light 
curves for two reasons.  First, as we discuss in greater detail below, the ejecta never overtakes the inner 
surface of the A00 shell because its radiation displaces the low-mass shell outward.  Second, the radiation 
front propagates to greater distances through the shell because of its lower density, so there is a larger 
separation between the front and the $\tau =$ 20 surface.  As a result, the SPECTRUM code may not 
resolve the radiating region of the flow and capture small-amplitude fluctuations in luminosity in diffuse 
shells as well as in more massive shells. In the A02 - A04 shells the radiation front and the $\tau =$ 20 
surface are much closer together and the radiating region is better resolved.

\subsection{Collision with the Shell}

\begin{figure*}
\begin{center}
\begin{tabular}{cc}
\epsfig{file=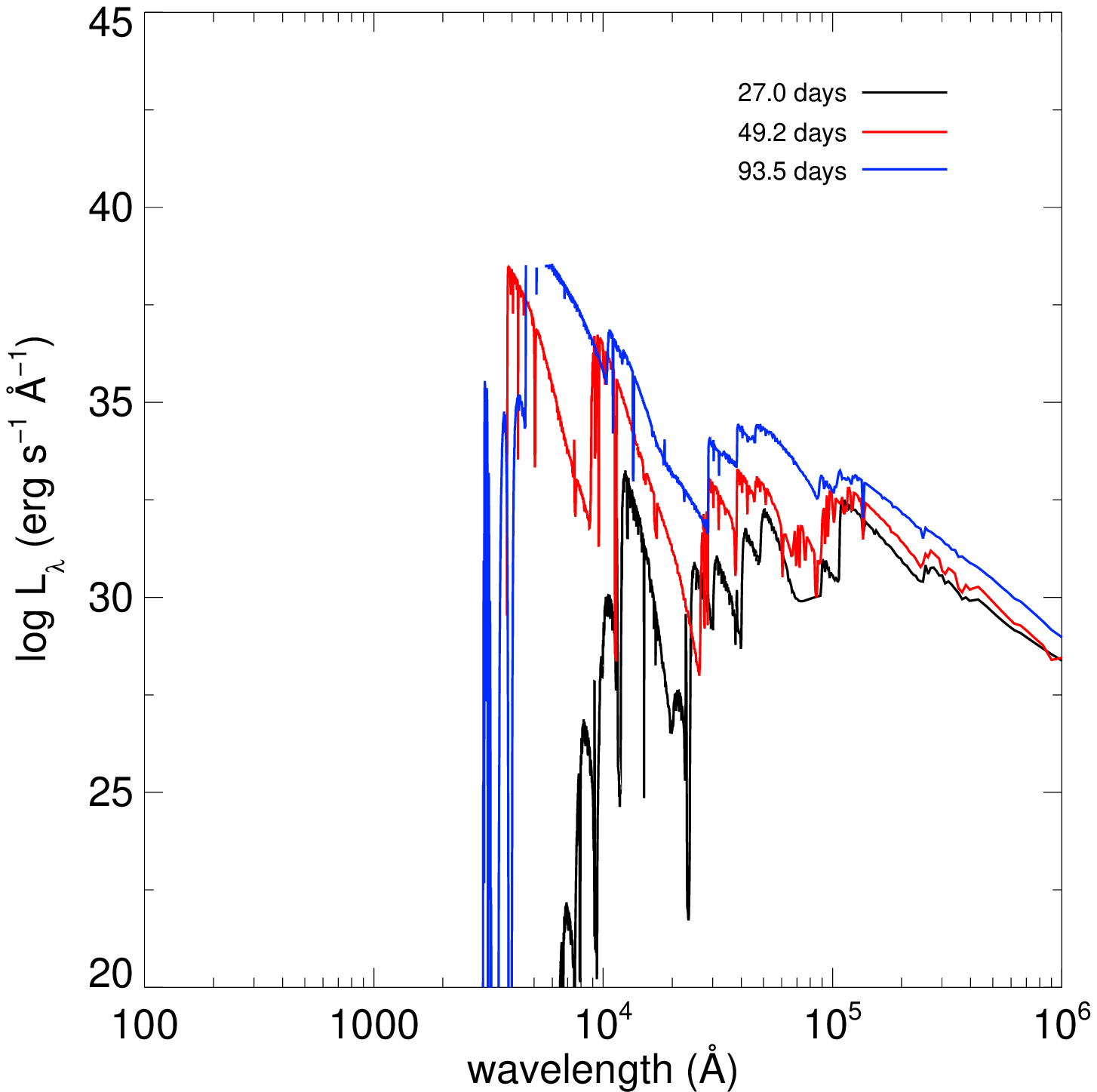,width=0.45\linewidth,clip=} & 
\epsfig{file=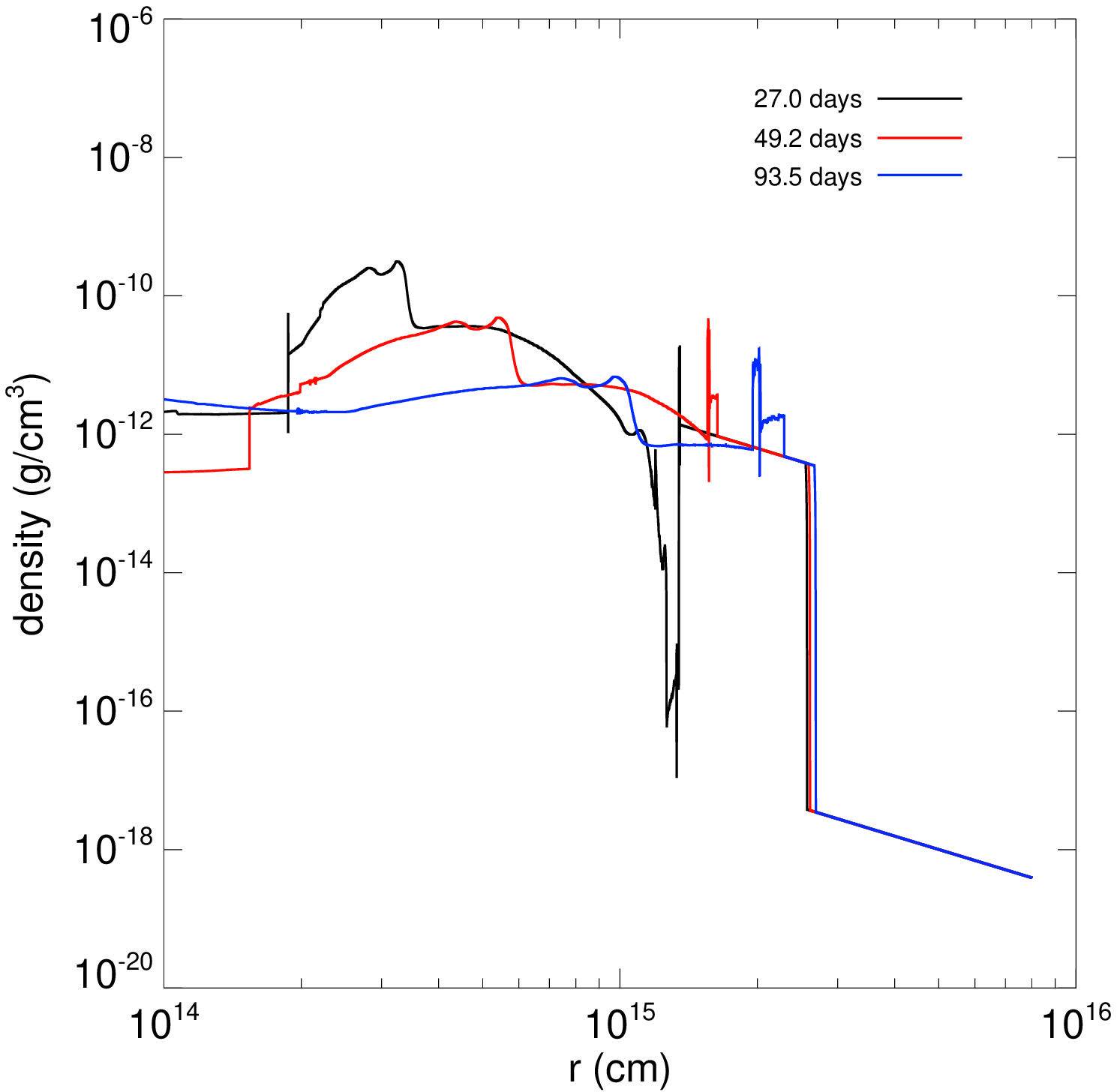,width=0.45\linewidth,clip=} \\
\epsfig{file=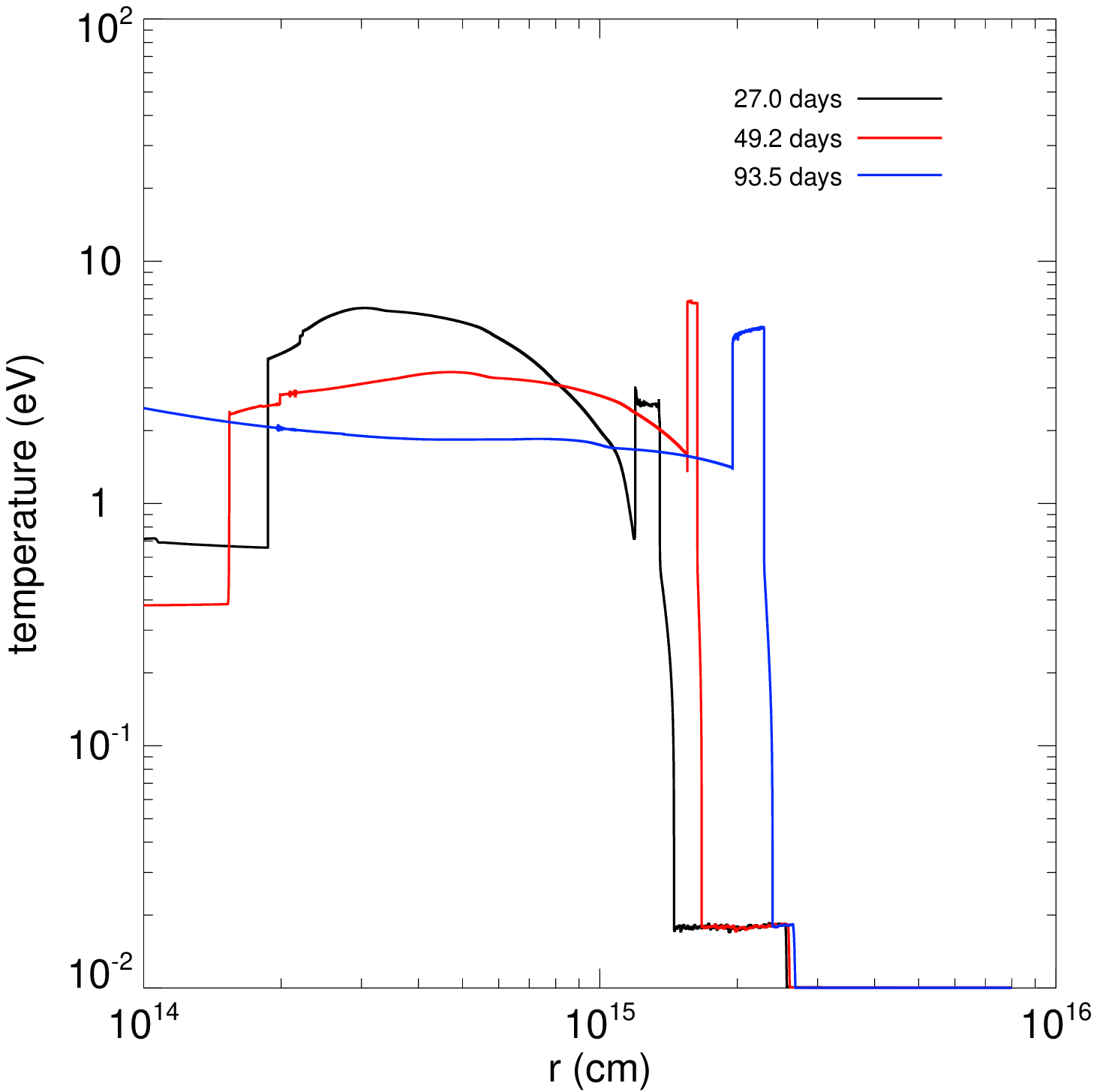,width=0.45\linewidth,clip=} &
\epsfig{file=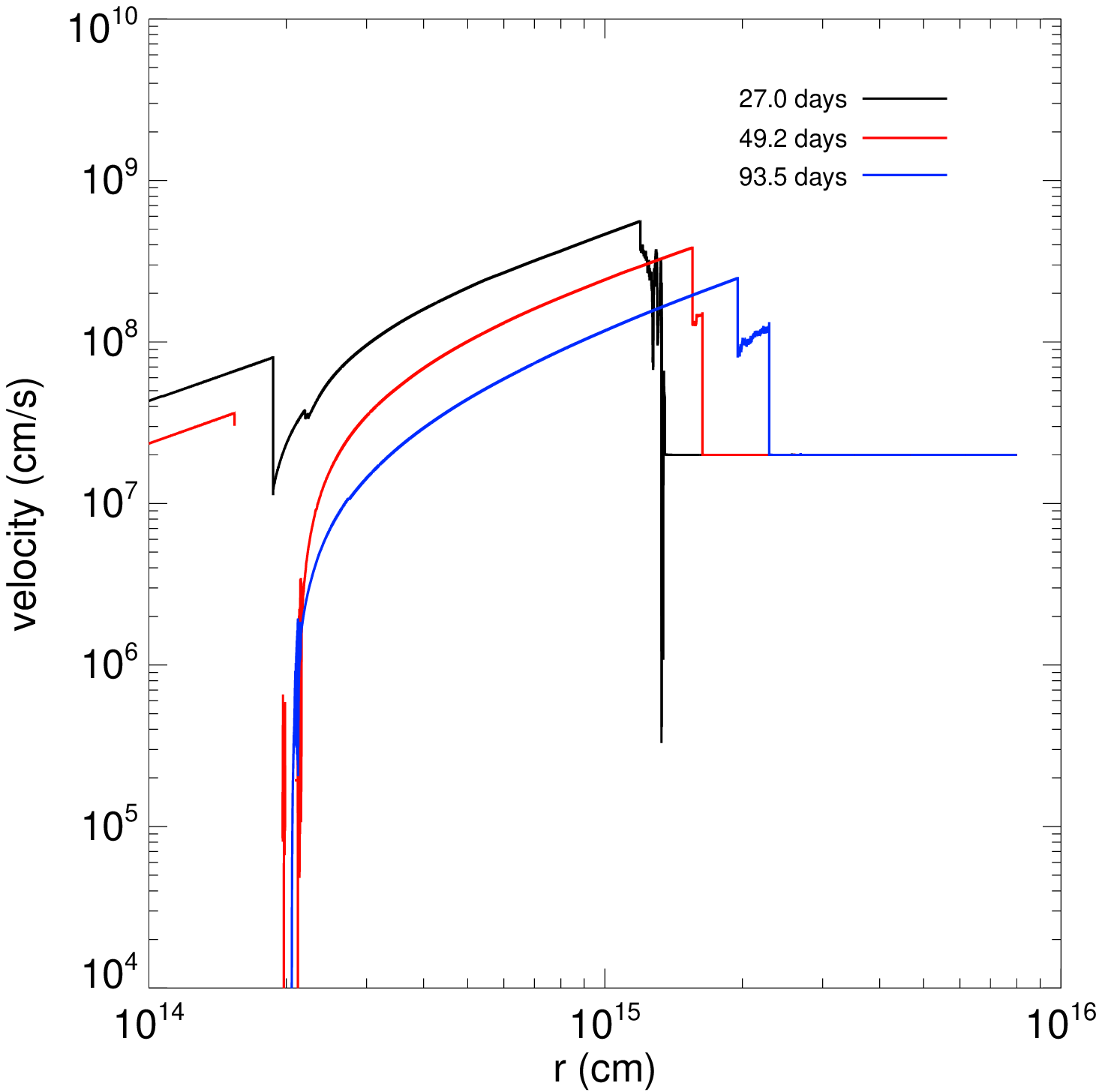,width=0.45\linewidth,clip=}
\end{tabular}
\end{center}
\caption{The collision of the ejecta with the A04 shell.  Upper left panel:  spectra; upper right panel: 
densities; lower left panel:  radiation temperatures; lower right panel:  gas velocities. The formation 
of a reverse shock as the ejecta plows through the shell can be seen in the double peak in the 
velocity profiles at 49.2 and 93.5 days.}
\label{fig:A04_2}
\end{figure*}

We show the ejecta just before and after its collision with the shell at 27 days and 49.2 days in Figure 
\ref{fig:A04_2}.  When it reaches the inner surface, the ejecta drives a strong shock into the shell, heating 
it to $\sim$ 7.5 eV.  The shock drives the inner surface outward, piling it up into the density spike visible at 
1.5 $\times$ 10$^{15}$ cm at 49.2 days.  This 7.5 eV spike is the source of the jump in luminosity at 32 
days in the left panel of Figure \ref{fig:lc}.  The magnitude of the jump depends on the density of the shell, 
with diffuse shells allowing more radiation to pass through them.  The light curves appear to plateau from 
32 - 60 days, but the luminosity rapidly rises as the shock propagates through the shell and there is less 
gas between the shock and the outer surface of the shell to absorb or scatter the photons, as we show at 
32 - 100 days in the right panel of Figure \ref{fig:lc}.  The rise is most pronounced in the densest shells 
because they initially attenuate the most radiation.

\begin{figure*}
\begin{center}
\begin{tabular}{cc}
\epsfig{file=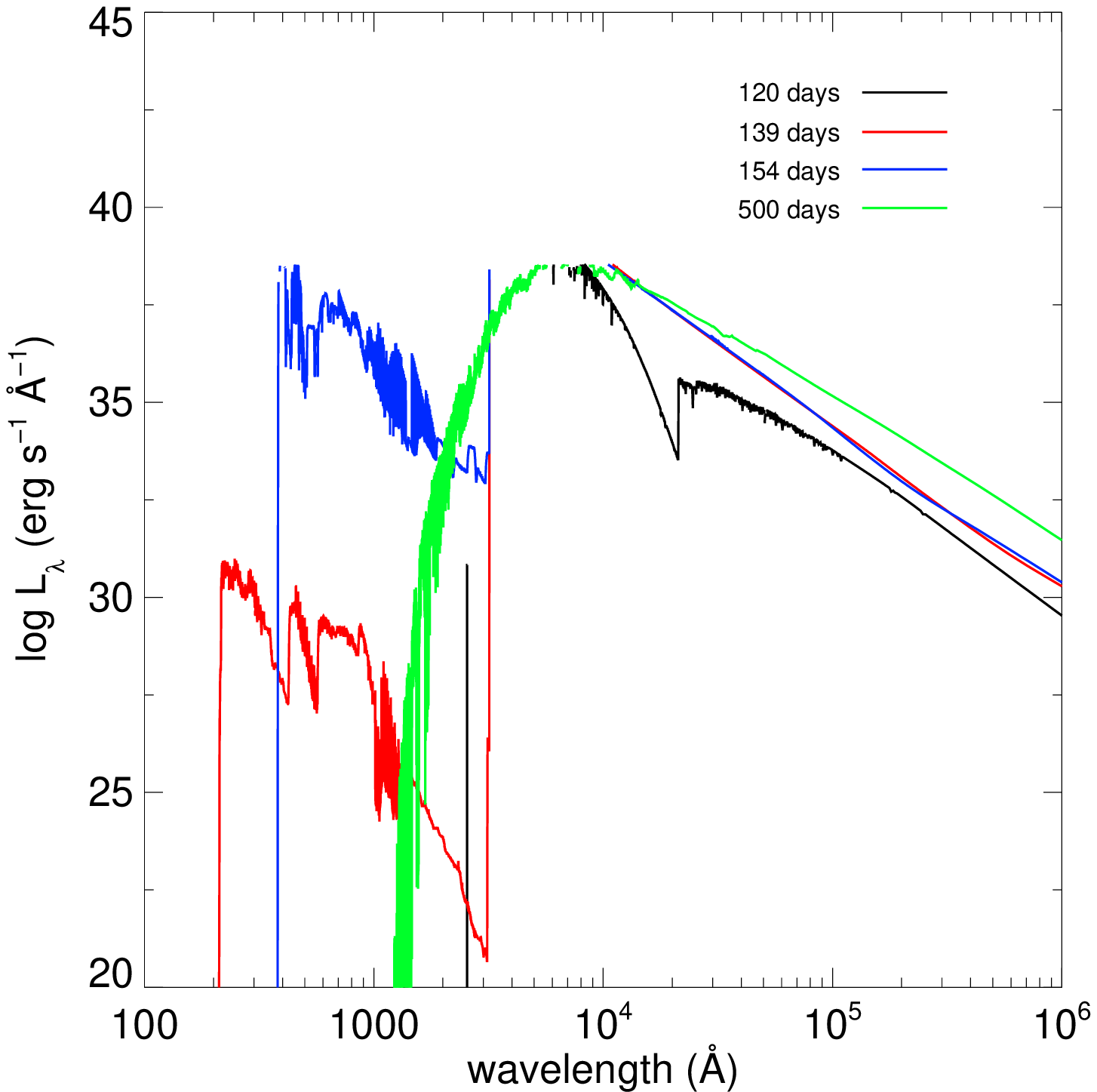,width=0.45\linewidth,clip=} & 
\epsfig{file=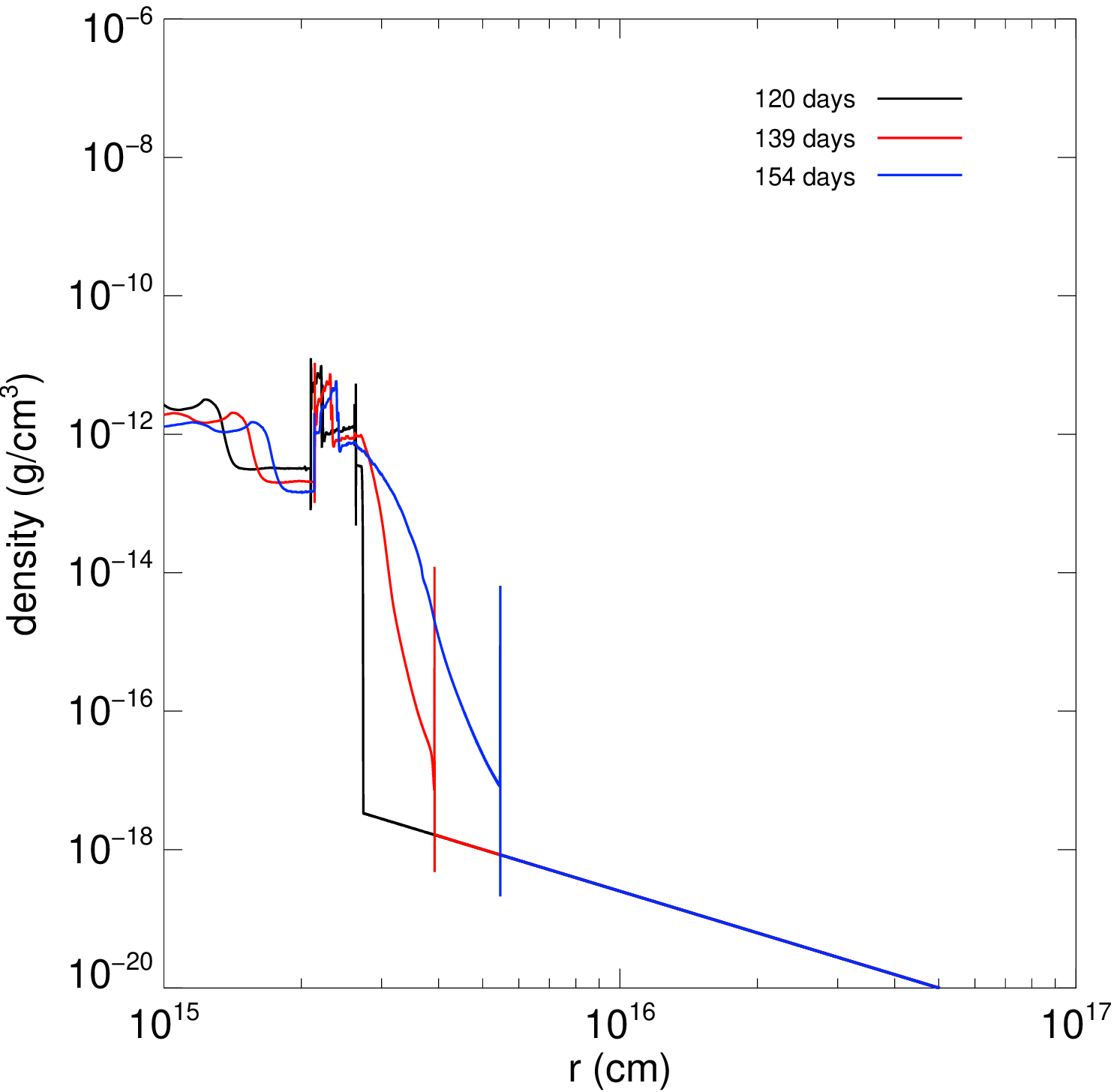,width=0.45\linewidth,clip=} \\
\epsfig{file=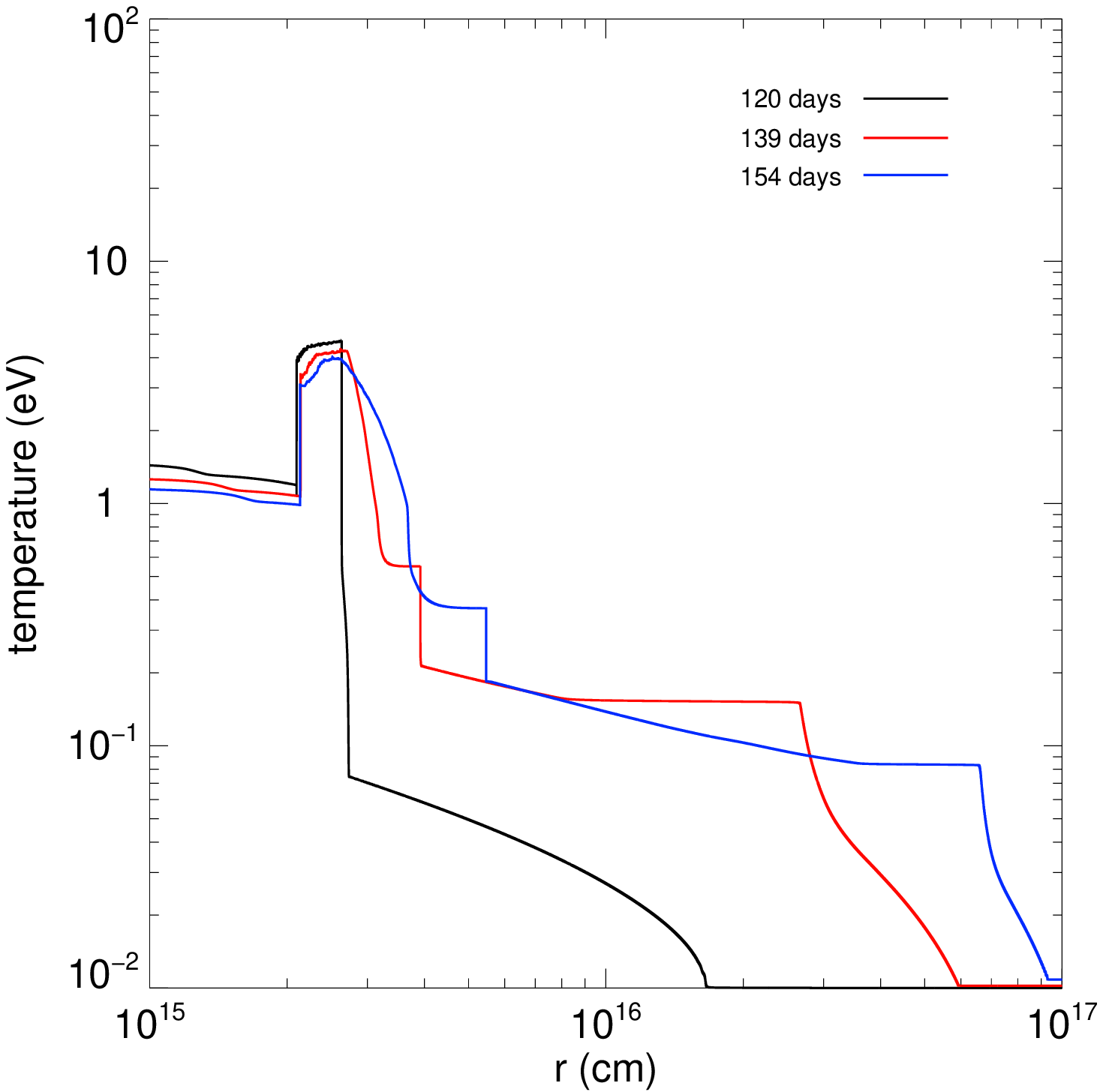,width=0.45\linewidth,clip=} &
\epsfig{file=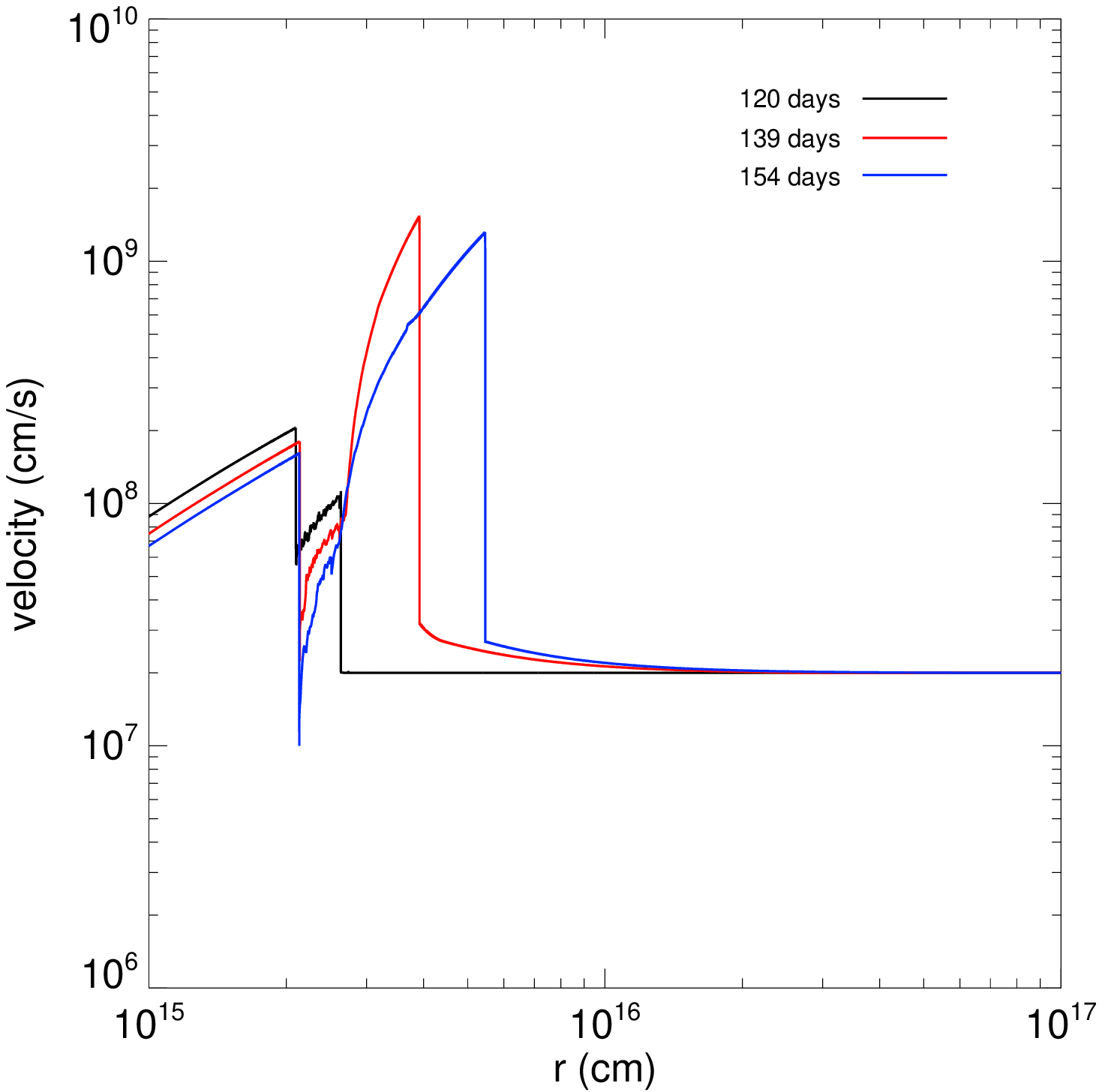,width=0.45\linewidth,clip=}
\end{tabular}
\end{center}
\caption{Shock breakout from the A04 shell. Upper left panel: spectra; upper right panel: densities; lower 
left panel:  radiation temperatures; lower right panel:  gas velocities.  The initial trapping and subsequent 
escape of radiation into the IGM can be seen in the absence of a plateau in radiation temperature ahead
of the shock at 129 days and its appearance and propagation at 139 and 154 days.  Breakout from the
shell is most evident in the jump in gas velocities between 120 and 139 days.}
\label{fig:A04_3}
\end{figure*}

As the ejecta plows up the shell a reverse shock forms and detaches from the forward shock, 
backstepping into the interior in the frame of the flow.  The reverse shock can be seen at 2.0 $\times$ 
10$^{15}$ cm in the density and temperature profiles at 93.5 days. As it recedes from the forward shock, 
the reverse shock loses pressure support because of radiative cooling by H and He lines in the shocked 
gas and it retreats back toward the forward shock.  However, as the forward shock continues to plow up 
the shell the reverse shock is again driven back into the interior.  The cyclical heating and cooling of 
postshock gas associated with the oscillation of the reverse shock causes the ripples in the luminosities 
from 32 - 100 days in the A02 - A04 light curves.  The period of oscillation is governed by cooling rates 
in the gas \citep{imam84} and is independent of shell mass, but the amplitude of oscillation is somewhat 
correlated with shell density \citep[see also section 4.1 of][]{anet97}.  Such features are also found in 
Lyman alpha emission by primordial SN remnants at later times as they sweep up neutral gas in halos 
on larger scales \citep[see Figure 11 in][]{wet08a}.  The stability of radiative shocks in astrophysical 
contexts has been well studied \citep[e.g.,][]{chev82}.  We note that the spectrum becomes harder from 
27 to 93.5 days even though the shock cools because there is less absorption by the shell as the shock 
plows through it.

\subsection{Breakout from the Shell}

The shock breaks free from the outer surface of the shell at times that depend on the mass of the shell, 
from $\sim$ 45 days in A01 to $\sim$ 120 days in A04.  As shown in the velocity profiles at 120 and 139 
days in Figure \ref{fig:A04_3}, the shock abruptly accelerates in the sharp density drop just outside the 
shell.  Photons from the shock also stream into the surrounding medium, creating the second jump in 
luminosity to $\sim$ 10$^{43.5} $erg s$^{-1}$ that persists for 50 - 100 days in the right panel of Figure 
\ref{fig:lc}.  Because the peaks coincide with shock breakout, they also occur at 45 - 120 days.  Diffuse 
shells result in broader peaks because they allow photons to escape from greater depths and earlier 
times in the shell.  As we show at 120 days in the temperature profile, low-energy photons begin to leak 
through the shell well before shock breakout at 139 days.  

Radiation from the shock again blows off the outer layers of the shell, creating the density peak at the 
edge of the radiation front at 4.0 $\times$ 10$^{15}$ cm at 139 days.  These photons ionize the 
envelope beyond the shell, allowing more energetic photons with $\lambda>$ 100 \AA\ to pass through, 
as shown in the spectrum at 154 days.  As the shock expands it cools, and the ambient density falls.  
From 154 to 500 days both of these effects are evident in the spectrum.  The peak of the spectrum 
shifts to longer wavelengths but more and more high energy photons escape into the IGM.  At 500 days 
the spectrum is nearly blackbody when its cutoff at high energies has fallen below the Lyman limit and 
ambient densities are very low.  From 250 - 500 days all five light curves gradually dim as the shock 
expands and cools.

\subsection{A00}

\begin{figure*}
\begin{center}
\begin{tabular}{cc}
\epsfig{file=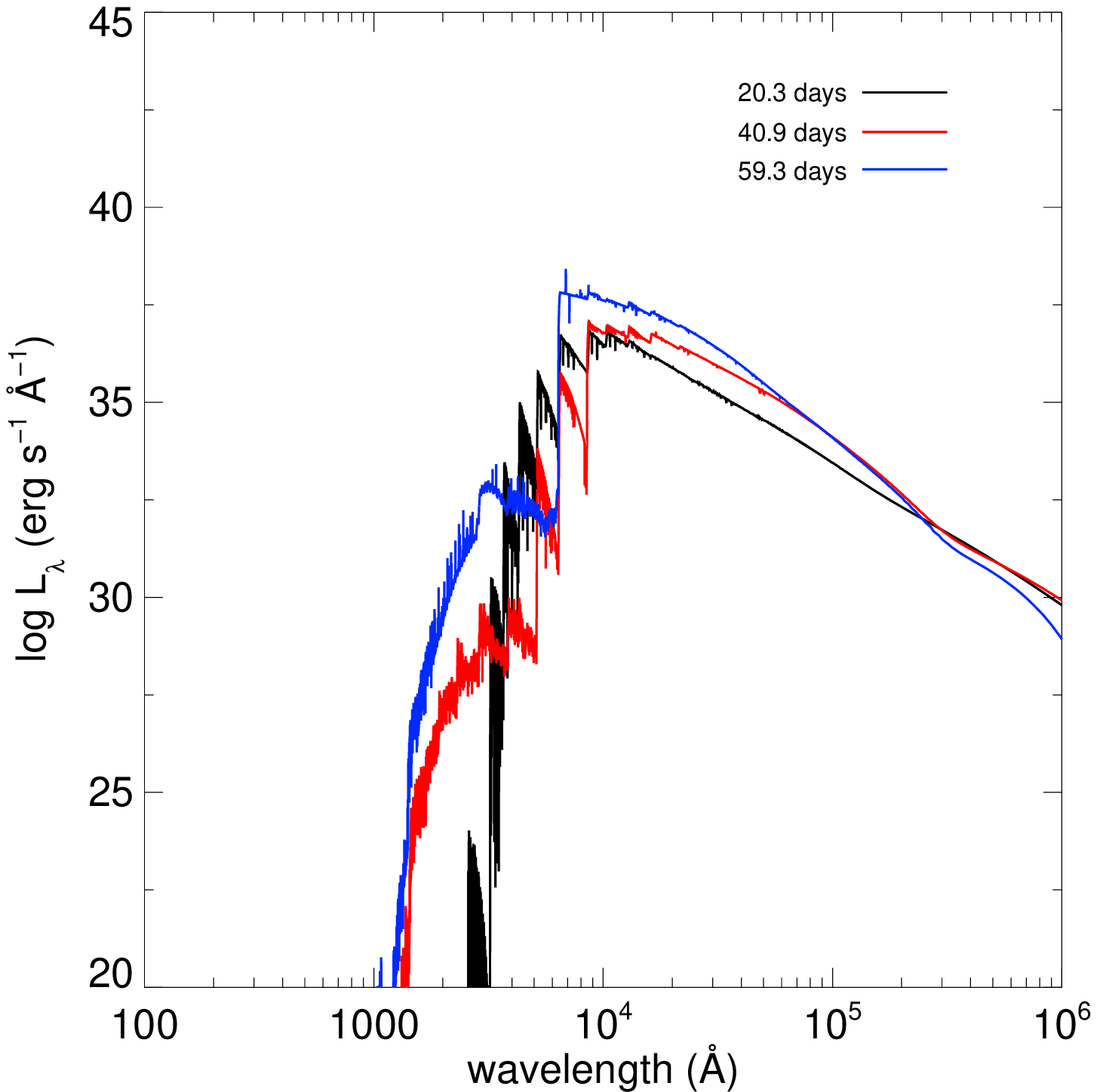,width=0.45\linewidth,clip=} & 
\epsfig{file=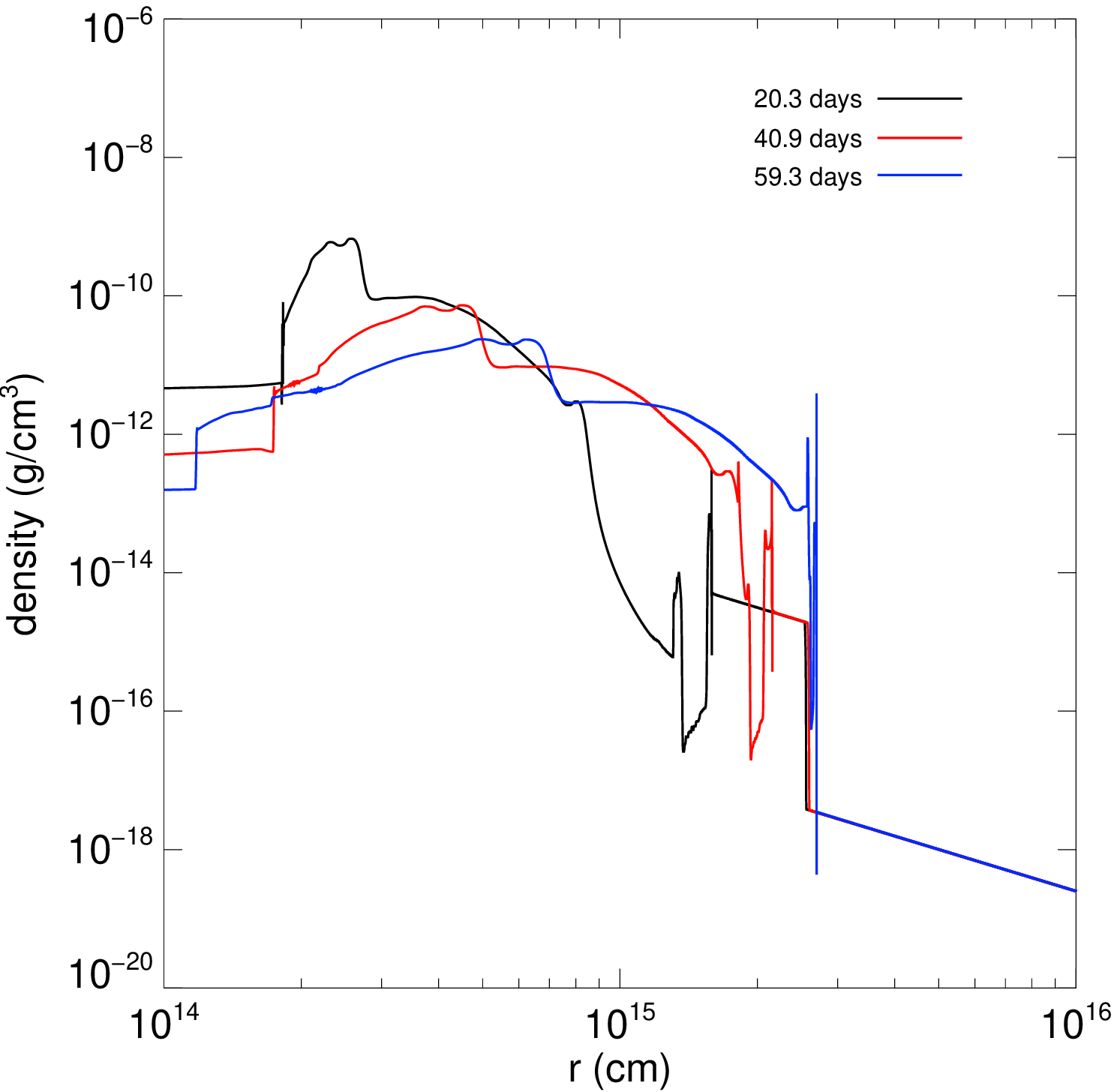,width=0.45\linewidth,clip=} \\
\epsfig{file=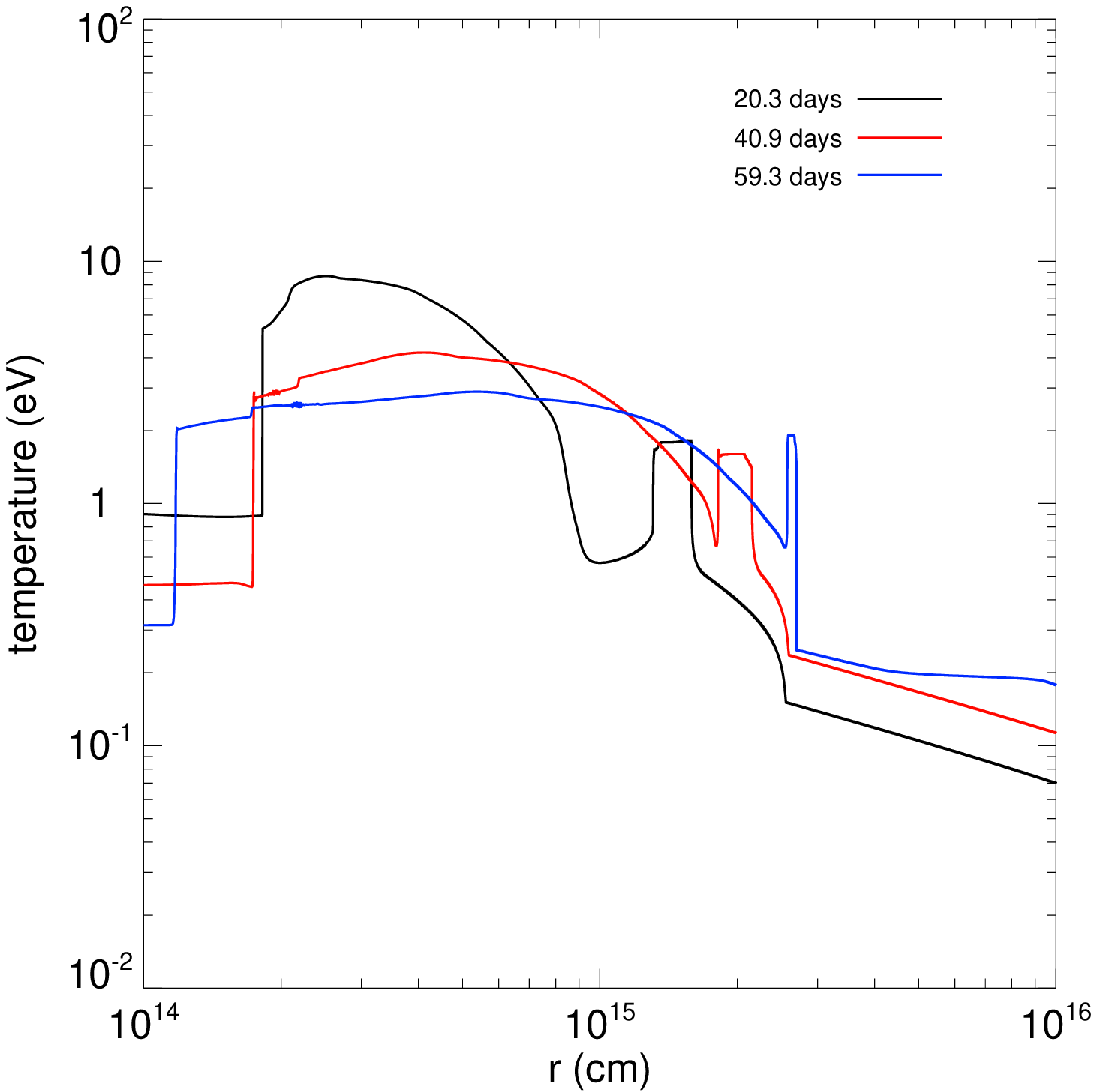,width=0.45\linewidth,clip=} &
\epsfig{file=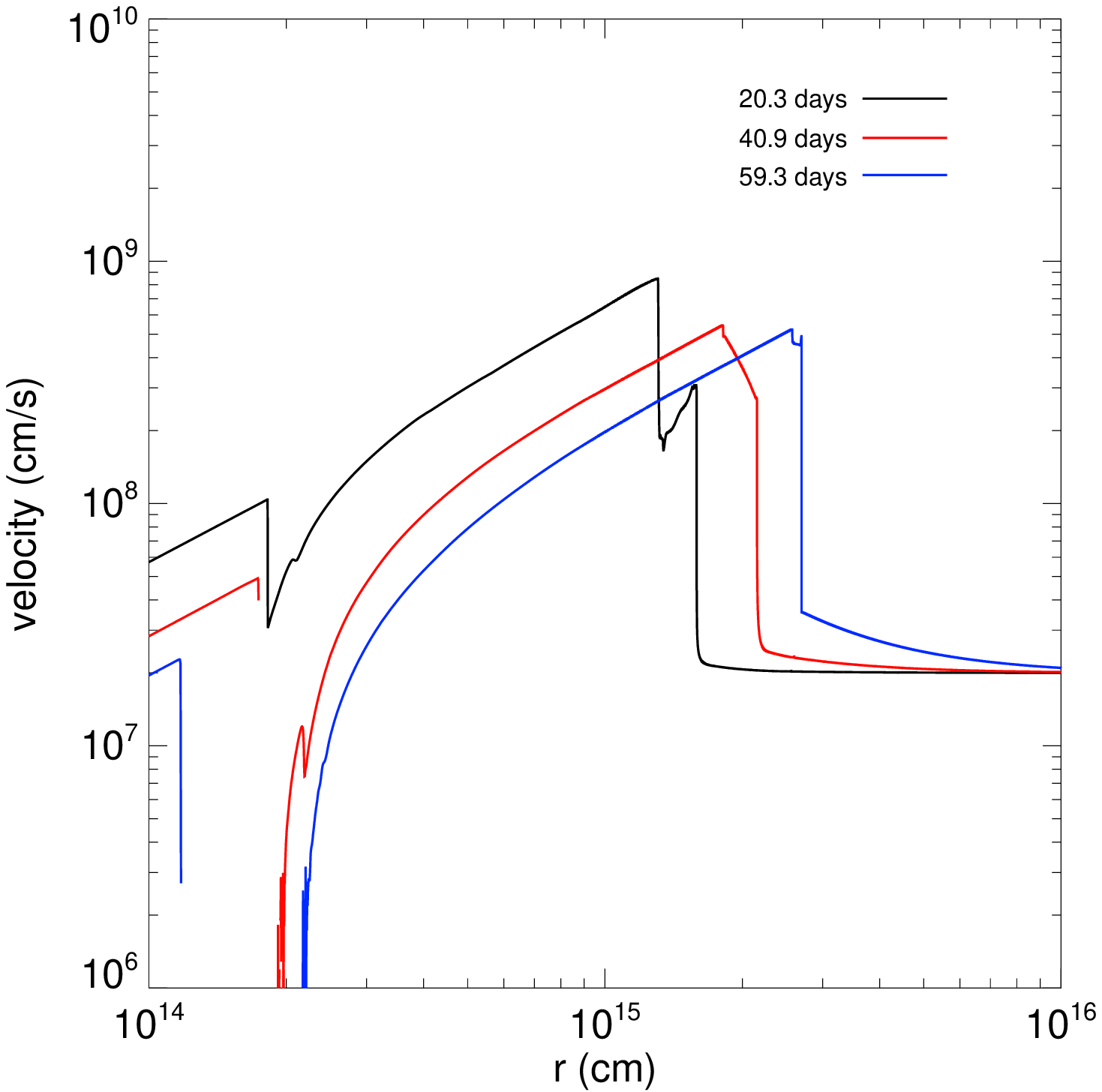,width=0.45\linewidth,clip=}
\end{tabular}
\end{center}
\caption{The collision of the ejecta with the A00 shell.  Upper left panel:  spectra; upper right panel:  
densities; lower left panel:  radiation temperatures; lower right panel:  gas velocities.  Note from the
profiles that the ejecta does not fully close the gap with the shell; radiation in the gap instead propels
the shell forward.}
\label{fig:A00}
\end{figure*}

The A00 explosion evolves somewhat differently because its shell is so diffuse.  The radiative 
precursor crosses the gap between the ejecta and the shell in $\sim$ 9 days, but radiation from 
the ejecta has already permeated and warmed the shell.  The precursor piles up in a thin layer at 
the inner surface of the shell as in the other cases, but as the ejecta approaches this surface the 
gap is never fully closed.  Radiation in the gap both displaces the inner surface outward and drives 
a reverse shock into the ejecta, as we show at 20.3, 40.9 and 59.3 days in Figure \ref{fig:A00}. The 
ejecta drives the inner surface forward while remaining separated from it by radiation forces.  As 
shown at 59.3 days, there is still a gap between the ejecta and the shell, and radiation from the 
shock has leaked through the shell.  By this time the outer edge of the shell has also begun to be 
displaced outward and it is quickly accelerated to the same velocity as the ejecta.  

This sequence of flow has several consequences for the A00 light curve.  First, as seen in Figure
\ref{fig:lc}, the luminosity from 8 - 32 days is higher because more radiation from the collision of the 
precursor with the shell gets through the shell.  Next, because the ejecta following the precursor is 
prevented from colliding with the shell by radiation in the gap between them, the rise in luminosity 
as the ejecta approaches the shell is very gradual, without the well-defined jump at $\sim$ 32 days 
in the A02 - A04 light curves.  Besides being more transparent to radiation, the A00 shell is also 
swept up more quickly than the others. The shock that is driven by the radiative gap reaches the 
outer surface of the shell at 60 days, when its luminosity peaks.  From 32 - 60 days the shock is far 
less luminous than in more massive shells because the A00 shell is being rapidly accelerated to the 
velocity of the ejecta.  Consequently, the kinetic energy of the ejecta is not efficiently converted into 
thermal energy, and the shock is weaker and dimmer while it is in the shell.  Finally, the luminosity 
jump at breakout in the other shells is absent in the A00 light curve because the ejecta never actually 
passes through the shell.  Supernovae in low-mass shells are much dimmer than those in shells 
above 1 \Ms.

The A00 light curve rebrightens between 110 and 220 days because the photosphere sinks into the 
ejecta and encounters layers of higher density and temperature, so the broad peak at 220 days is 
due purely to optical depth.  A similar feature is visible at 240 days in the A01 light curve.  The 
evolution of the A01 explosion is intermediate to that of A00 and the others.  Its luminosity gradually 
rises from 8 - 32 days because radiation from the impact of the outer layers of the star with the shell 
gets through the shell and intensifies as gas piles up at the inner surface. Next, because a radiatively 
supported gap again softens the collision of the ejecta with the shell, the rise in luminosity upon impact 
is again more gradual than with more massive shells but more prominent than in the A00 light curve.  
For shell masses greater than 1 \Ms, breakout luminosities from the shell are 10$^{43.5}$ - 10$^{43.8
}$ erg s$^{-1}$.  

\subsection{Earlier Simulations}

\begin{figure*}
\begin{center}
\begin{tabular}{cc}
\epsfig{file=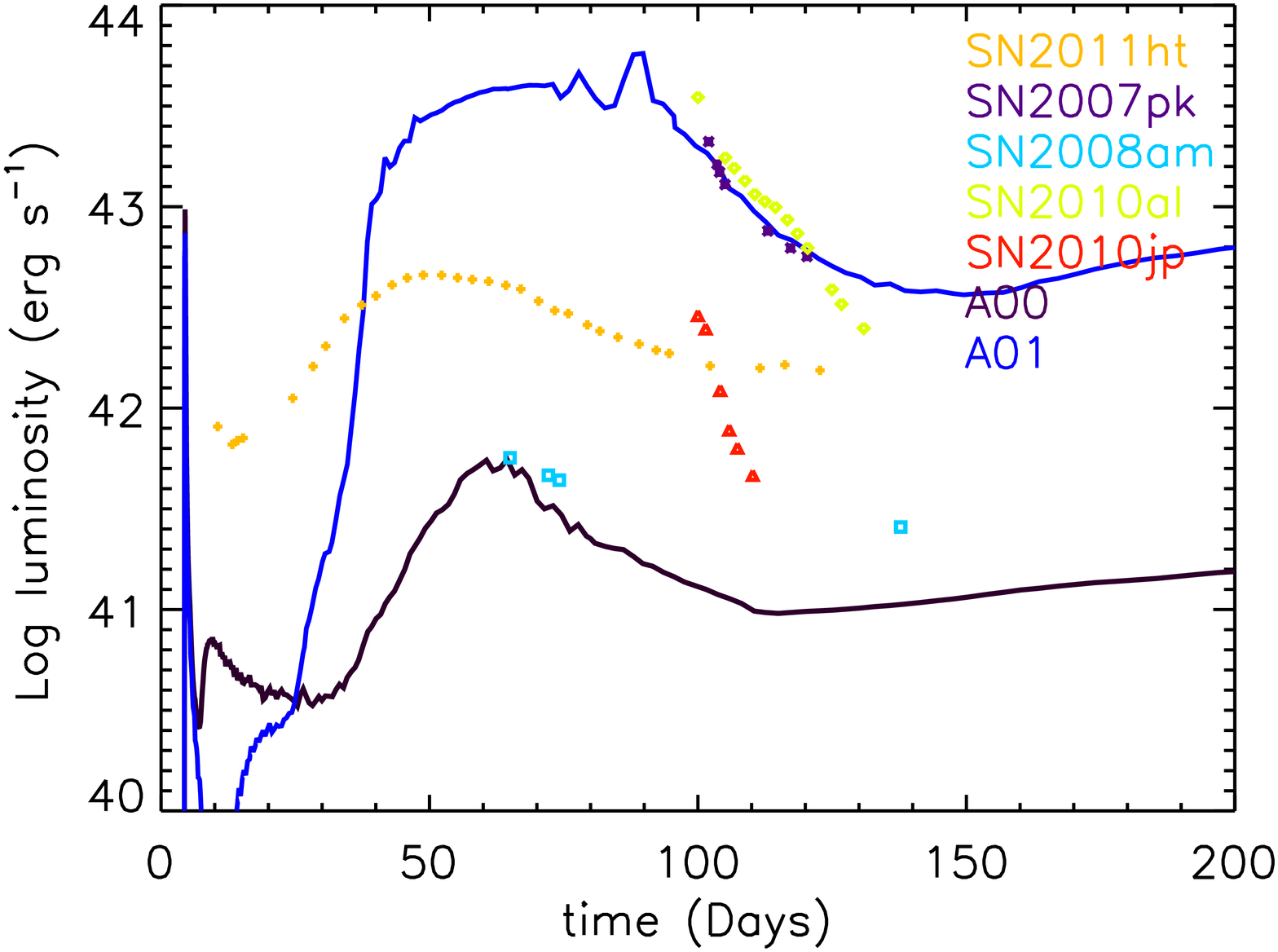,width=0.45\linewidth,clip=} &
\epsfig{file=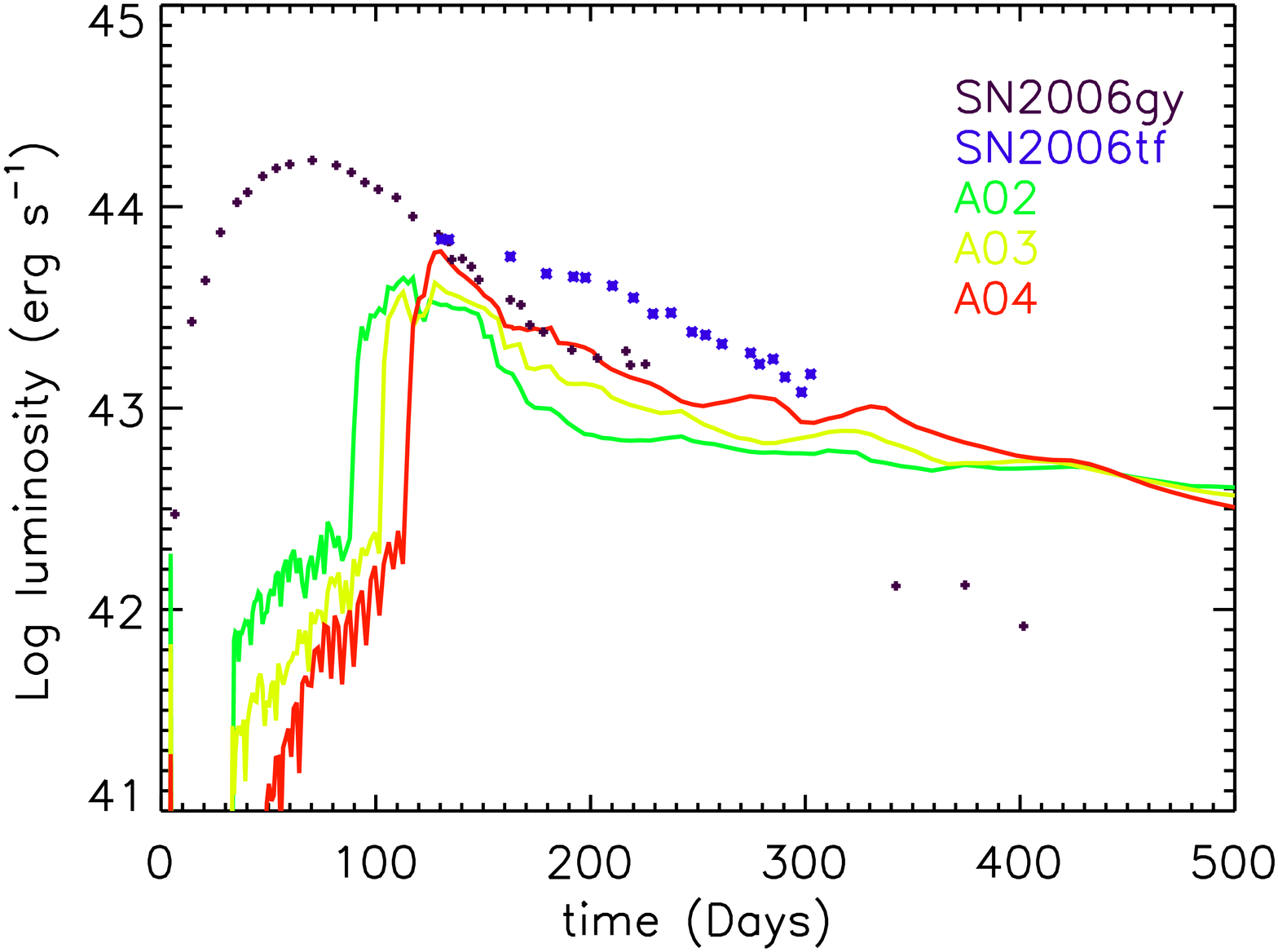,width=0.45\linewidth,clip=} \\
\end{tabular}
\end{center}
\caption{Bolometric light curves for Type IIn SN candidates in the local universe overlaid with the A00 
- A04 light curves.  Left: less luminous Type IIne and the A00 and A01 models.  Right: superluminous
Type IIn candidates with the A02 - A04 models.  In the observations, there is a hint of shock breakout
from the surface of the star with SN2011ht at 10 days that is absent in the other data.}
\label{fig:observe}
\end{figure*}

Not surprisingly, radiation transport leads to important qualitative differences between our light curves 
and those of \citet{vmarle10}, even though we use the same shells in our models.  \citet{vmarle10} 
equate energy losses due to optically thin radiative cooling in the gas with the bolometric luminosity of 
the SN remnant and hence do not account for absorption by the optically thick shell.  These losses are 
greatest when the gas is most strongly shocked, when the ejecta strikes the inner surface of the shell.  
Consequently, the \citet{vmarle10} light curves peak upon initial collision when in reality most of these 
photons are absorbed in the lower layers of the shell (compare their Figure 11 at 25 days to the right 
panel of Figure \ref{fig:lc} at 32 days).  On the other hand, because collisional gas cooling scales as 
$\rho^2$ in \citet{vmarle10}, their bolometric luminosities plummet when the shock breaks out into the 
low-density medium beyond the shell (compare Figure 11 at 60 - 100 days to Figure \ref{fig:lc} at 50 - 
100 days).  As shown in Figure \ref{fig:lc}, the light curve actually peaks at this point because photons 
previously trapped in the shock are free to stream into the IGM, and as the shock rushes down the 
density gradient it heats and radiates additional energy.  When the SN collides with the shell our light 
curves are therefore initially dim and later peak upon breakout from the shell, while the converse is true 
of the \citet{vmarle10} light curves.  We note that our shock propagates through the shells in about the
same times as in \citet{vmarle10}.

Interestingly, both approaches yield similar widths for the light curves at peak luminosity.  The width 
of the peak is closely tied to the propagation time through the shell in \citet{vmarle10} but is largely a 
function of the subsequent expansion and cooling of the remnant in our models.  Since shock 
temperatures peak at only $\sim$ 8 eV upon impact with the shell, we do not find any x-ray emission 
after breakout from the surface of the star but the SN is bluer out to much later times than in z40G by 
itself.  This property makes Type IIn SNe ideal candidates for high redshift detection.  We note that 
some of the light curves in Figures 10 and 11 of \citet{vmarle10} exhibit the same flickering as in ours, 
probably also because of oscillations of reverse shocks due to radiative cooling.  They are especially 
prominent as the shock plows through the A00 shell in Figure 11, when a reverse shock is likely to 
detach from the forward shock.  Finally, because \citet{vmarle10} model the SN profile as a free 
expansion and do not consider breakout from the star, none of the features from 8 - 32 days in our 
light curves are present in theirs.

We note that while the shock never reaches temperatures at which it would emit x-rays in our models,
this in part is due to the fact that we assume that ions and electrons are closely coupled.  Because the 
electrons and the photons are also closely coupled, the ions rapidly lose energy to the radiation field.
Because $e_{gas} = C_VT$ and $e_{rad} = aT^4$, large transfers of energy into the radiation field do
not result in large changes in its temperature, so the radiation and matter temperatures remain modest 
and there are only mild deviations between them in our simulations.  In reality, the ions may not couple 
efficiently to electrons when the shock collides with the inner surface of the shell or breaks free from its 
outer surface and the shock may heat them to much higher temperatures than the radiation temperature. 
Under these circumstances the ions may drive a small fraction of the electrons out of equilibrium with the 
radiation field and cause them to emit x-rays in real SN remnants.  

Comparison of the more recent light curves of \citet{moriya12} with ours is more problematic, in part 
because they assume much higher explosion energies (10 - 50 $\times$ 10$^{51}$ erg, rivaling those 
of pair-instability SNe) and because they adopt a different structure for the shell. As a result, their peak 
luminosities are 5 - 10 times higher than ours but have similar widths.  Our bolometric light curves are 
otherwise qualitatively similar to theirs (compare their Figure 5 to the left panel of Figure \ref{fig:lc}).  
Their large explosion energies are necessary to achieve the peak bolometric luminosities of SN 2006gy 
\citep{nsmith07b,nsmith08} and they are not normal core-collapse events.

\section{Type IIn SNe in the Local Universe}

\begin{figure*}
\begin{center}
\begin{tabular}{cc}
\epsfig{file=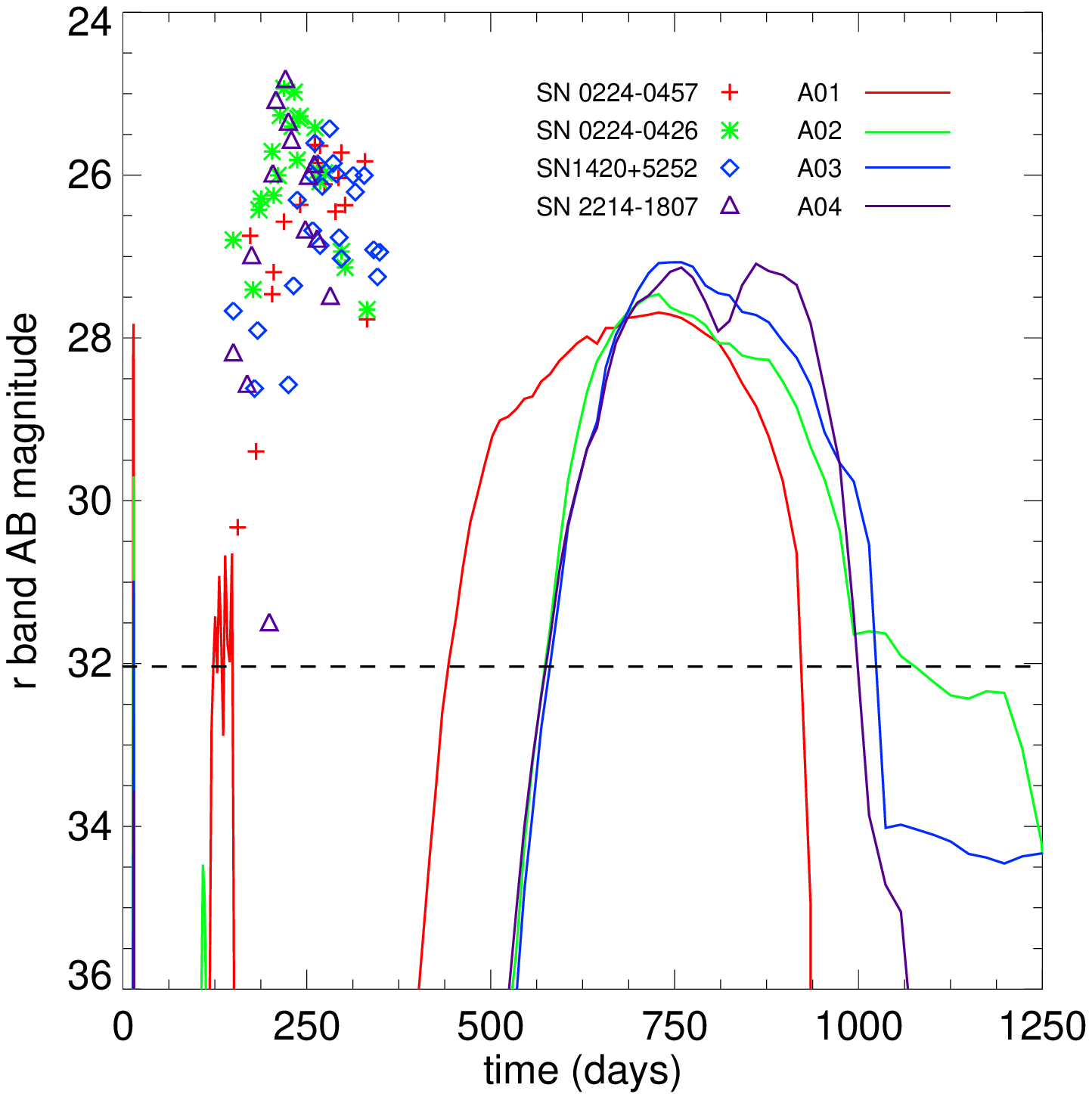,width=0.45\linewidth,clip=} &
\epsfig{file=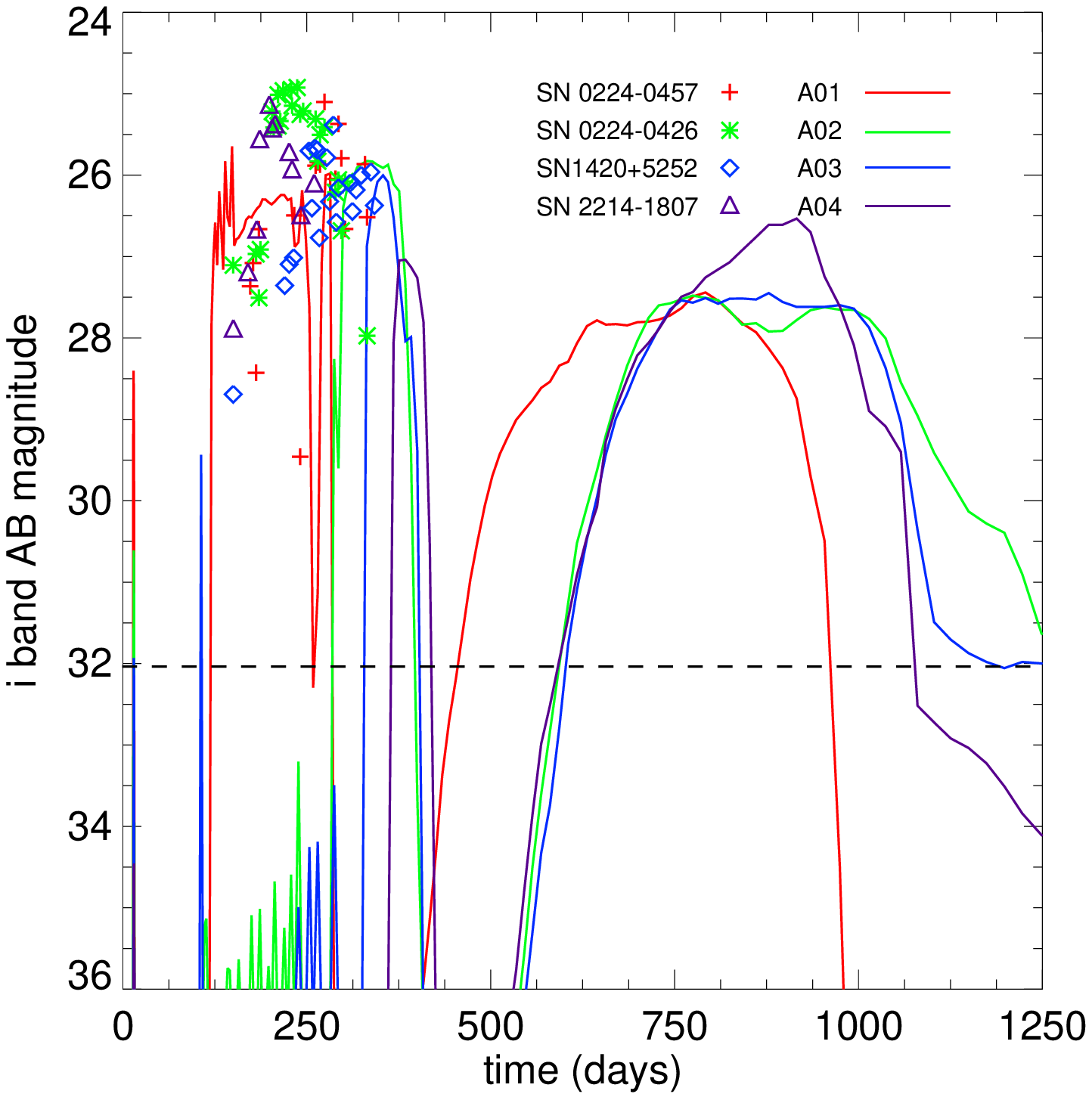,width=0.45\linewidth,clip=} \\
\end{tabular}
\end{center}
\caption{Observations of $z \sim$ 2.2 Type IIn SN candidates vs. simulations.  Left: r band. Right: i 
band.  In both panels, the first peak is due to the collision of the ejecta with the inner surface of the
shell and the second peak is due to shock breakout from the shell.  The first peak is mostly absent
in the r band because the photons are blueward of the 3000 \AA\ absorption feature of the spectra
in Figure \ref{fig:atten} and are cut off by the shell.}
\label{fig:cooke}
\end{figure*}

We now compare bolometric luminosities for our five explosions to those of Type IIn SNe discovered 
in the local universe:  SN2006gy \citep{nsmith07b,nsmith08}, SN2006tf \citep{nsmith08c}, SN2007pk 
\citep{prit12}, SN2008am \citep{chatz11}, SN2010al \citep{cooke10}, SN2010jp \citep{nsmith12}, and 
SN2011ht \citep{rom12,hmph12}.  SN2006gy and SN2006tf are the most luminous explosions in this 
class while the others exhibit more typical peak luminosities.  We group luminosities for A00 and A01 
with those of normal Type IIn SNe in the left panel of Figure \ref{fig:observe} \citep[see][for additional 
examples of Type IIn SN light curves]{kiewe12} and light curves for A02, A03 and A04 with those for 
the superluminous SN2006gy and SN2006tf in the right panel of Figure \ref{fig:observe}.  The rise of
the light curve is evident in SN2006gy and SN2011ht; the other five datasets only show its decline, 
making it difficult to pinpoint their explosion times.  We therefore assign them times that best align 
them with our simulations.  The observations of the normal brightness Type IIn SNe either have very 
few data points (SN2008am) or cover only short periods of time (SN2007pk, SN2010al, and SN2010jp).  

The central engines of Type IIn SNe, the core-collapse explosion, do not vary much with metallicity 
\citep{cl04,wh07} \citep[note also Figure 1 of][]{wf12}.  It is primarily the structure of the shell and 
its cooling properties that differentiate Pop III Type IIn supernovae from those today.  Fine-structure 
cooling by metals flattens shells into colder, thinner and denser structures than do H and He lines.  
The thickness of the shell governs the width of the main luminosity peak, so Type IIn events today 
would likely exhibit narrower peaks and sharper declines in bolometric luminosity, as with SN2010jp 
and SN2010al.  On the other hand, inefficient gas cooling keeps the shock hot and bright, with much 
slower decays in luminosity at later times like those in Figure \ref{fig:observe}.  More realistic 
treatments of the ejection would impose additional features on both zero-metallicity and enriched 
shells that are not present in our models.  Slow outbursts are usually preceded and followed by 
much faster winds.  The wind in front of the shell detaches from and races ahead of it, creating a 
rarefaction zone, while the wind behind the shell piles up at its inner surface and forms a hot 
termination shock \citep[see][]{met12a}. These structures will imprint additional features on the light 
curves that have not been captured by any simulations to date.  Type IIn SNe in more realistic shells 
in the local universe will be pursued in future simulations.
  
In spite of these limitations, the A00 and A01 light curves are consistent with those of the five less 
luminous Type IIne.  They fall between A00 and A01, with rates of decline that are similar to those 
in the simulations.  In particular, A01 is an excellent match to SN2007pk.  Because SN2011ht was 
detected during its rise and observed for longer times, it places tighter constraints on our models.  
Its luminosity peak falls almost directly between those of A00 and A01 on a log scale, and it has   
about the same width.  However, the ratio of peak to plateau luminosities is smaller for SN2011ht 
than for the two simulations.  The z40G explosion in a shell with a mass of $\sim$ 0.4 \Ms\ would 
yield the best match to SN2011ht.

SN2006gy, another Type IIn SN that was observed during its rise to peak luminosity, is nearly a 
factor of 3 brighter than A04, our brightest light curve.  As noted earlier, bolometric luminosities 
rose only slightly as the mass of the shell went from 6 to 20 \Ms, so more massive shells will not 
yield better agreement with SN2006gy. The dip at the beginning of the SN2006gy light curve may 
be photons from shock breakout from the surface of the star filtering through the shell, which only 
happens with less massive shells.  Taken together, these two facts suggest that a more powerful 
SN is needed to explain SN2006gy, not a more massive shell \citep[which is consistent with][]{
moriya12}.  SN2006tf, the other superluminous explosion, is only marginally brighter than A04. Its
bolometric luminosities are close to those of A04 at early and late times, but A04 falls more rapidly 
and then enters a plateau while SN2006tf declines more steadily.  As mentioned above, the 
plateau is likely due to inefficient H and He cooling in the A04 shell and would probably disappear 
if the gas was enriched with metals.  The A04 light curve in \citet{vmarle10} is in basic agreement
with SN2006gy for shell masses of 20-24 M$_\odot$ and wind velocities of 190 km s$^{-1}$.

Our light curves are in general agreement with recent Type IIn SNe, and some observations match 
our simulations extremely well.  Nevertheless, we do not expect exact agreement because we only
used one explosion and shell, and varied only the density of the shell.  The shells in our models are 
also somewhat different from those in real explosions.  From Figure \ref{fig:lc} it is clear that many 
of the properties of the shell can be extracted from the light curves, so they can be powerful probes 
of the circumstellar environment of the explosion.  

\section{z $\sim$ 2 Type IIn SNe}

In Figure \ref{fig:cooke} we compare r-band and i-band light curves for the A01 - A04 runs with 
those of Type IIn SNe recently discovered at 1.9 $< z < $ 2.4: SN 0224-0457, SN 0224-0426, SN 
1420$+$5252, and SN 2214$-$1807 \citep{cooke09,cooke12} with the Low Resolution Imaging 
Spectrometer \citep{lris,steid04,lrisb,rock10}.  These SNe reach peak AB magnitudes of  24.5 - 
26 and are visible for 200 -- 250 days in both bands.  Our simulated r-band light curves exhibit 
two peaks:  a brief initial peak lasting no more than 50 days with magnitudes below 28, and a 
second brighter and longer peak that reaches magnitudes of 27 -- 28 and lasts up to 500 days.  
The i-band light curves also exhibit two peaks but the first one is longer and brighter than in the 
r-band:  50 - 150 days with magnitudes of 26.5 to 27.5. In both bands the first peak is due to the 
collision of the SN ejecta with the inner edge of the shell and the second peak is due to shock 
breakout from the shell, when its photons are suddenly able to stream freely in the low density 
wind.  The first peak in the i band becomes brighter and longer as the shell mass decreases.  In 
Figure \ref{fig:A04_2}, the shock reaches temperatures of $\sim$ 10 eV and becomes quite 
luminous upon its collision with and initial advance into the shell.  However, these photons must 
filter through the shell, and the number that escape depends strongly on wavelength.  As shown 
in Figure \ref{fig:A04_2}, the opacity of the shell imposes a sharp cutoff on the spectrum at $\sim$ 
3000 \AA.  This explains the prominence of the first peak in the i band.  The photons that are 
redshifted into the i band from $z =$ 2.2 originate from the brightest region of the spectrum just 
redward of the 3000 \AA\ cutoff.  There is little luminosity in the r band because they are blueward 
of the cutoff and absorbed by the shell.

\begin{figure*}
\begin{center}
\begin{tabular}{cc}
\epsfig{file=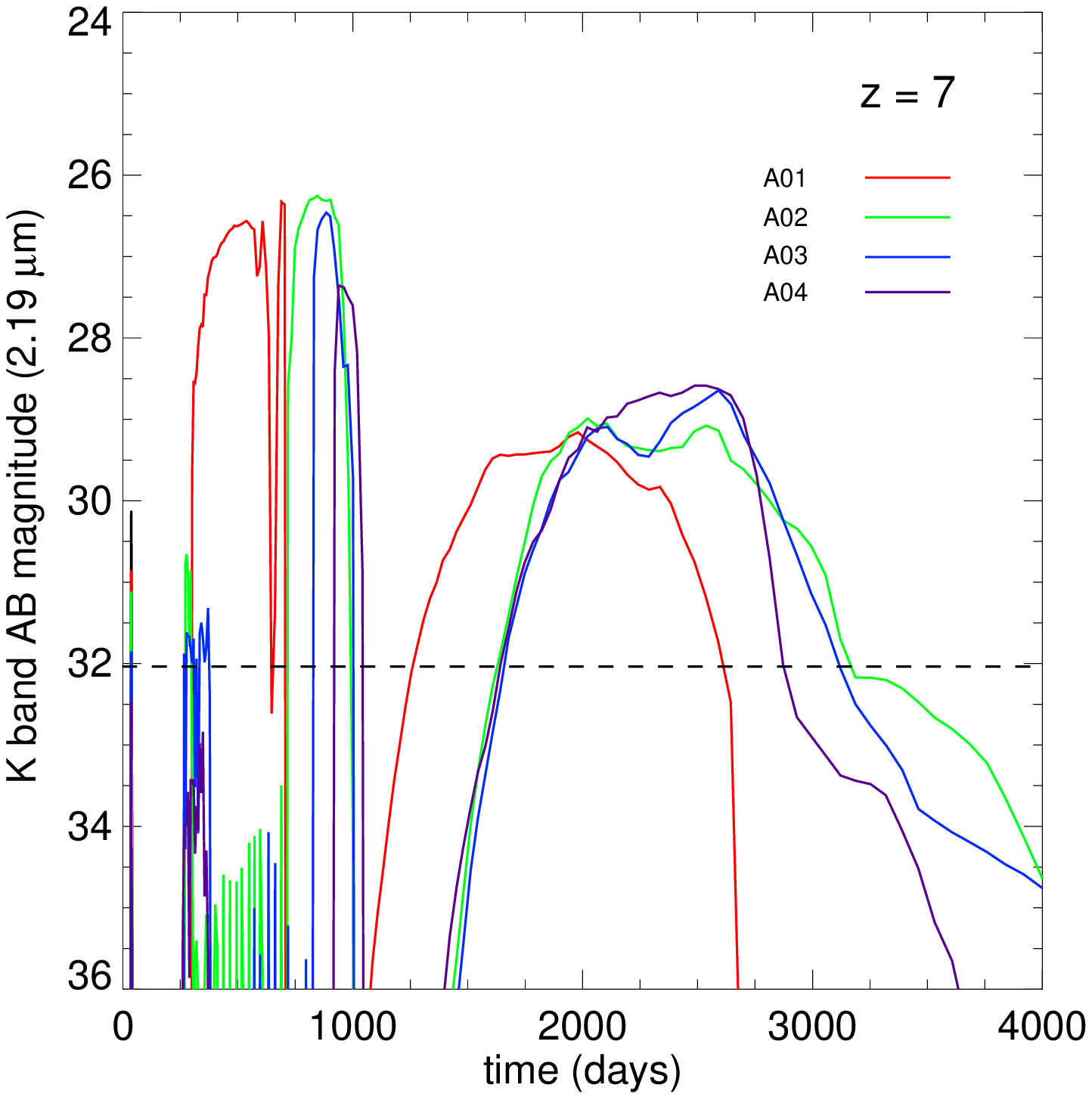,width=0.45\linewidth,clip=} & 
\epsfig{file=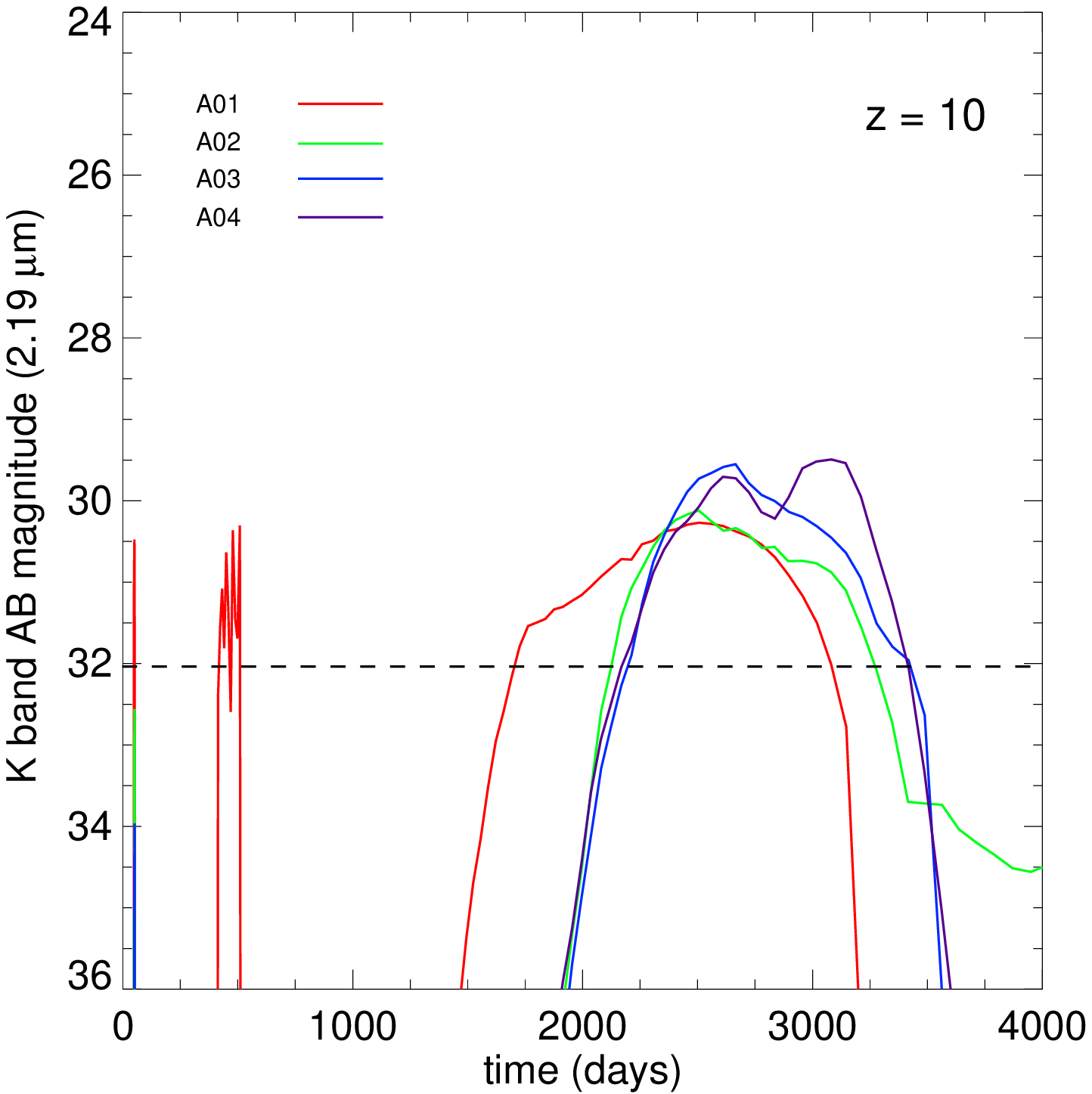,width=0.45\linewidth,clip=} \\
\epsfig{file=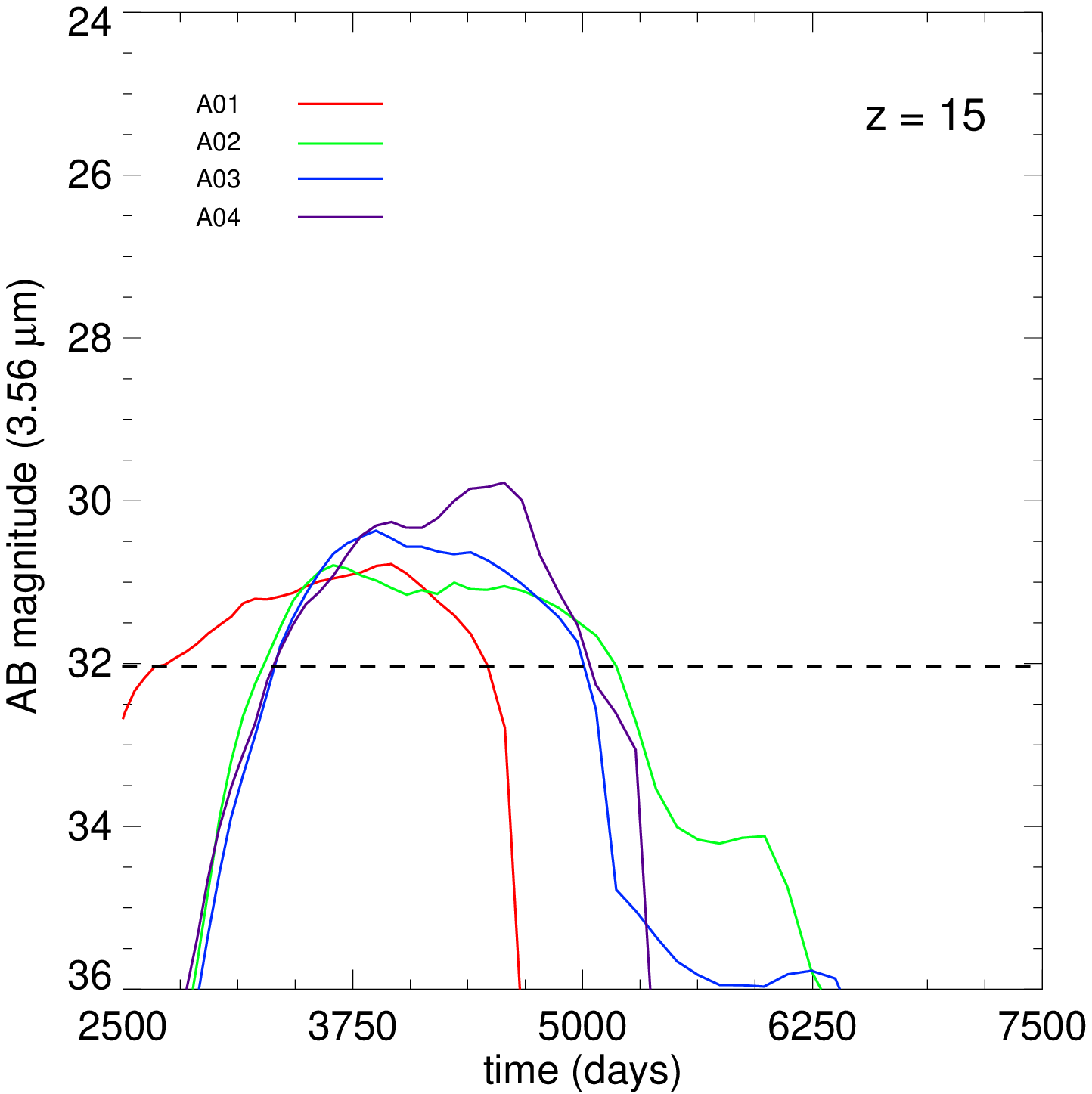,width=0.45\linewidth,clip=} &
\epsfig{file=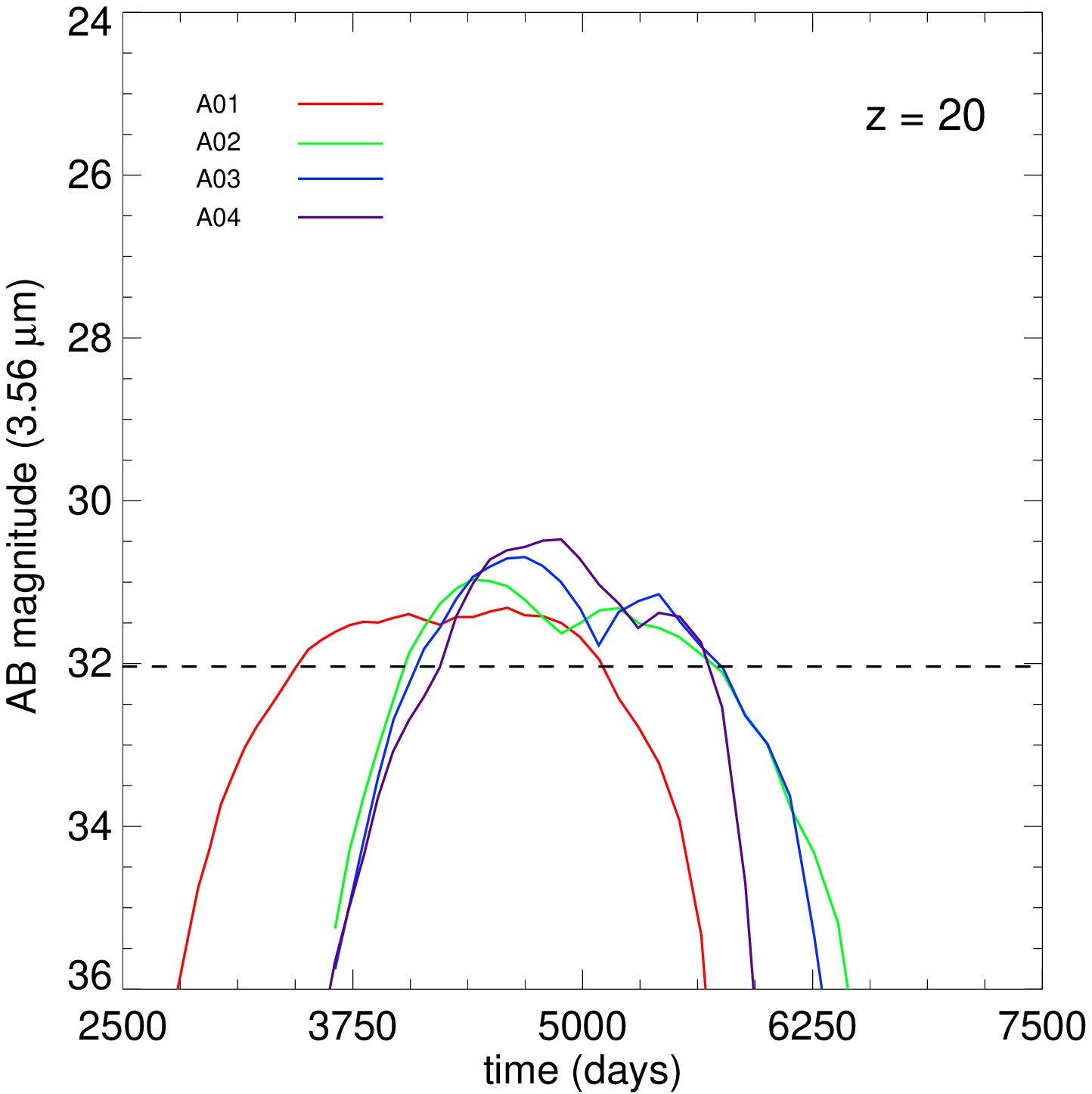,width=0.45\linewidth,clip=}
\end{tabular}
\end{center}
\caption{NIR light curves for all five shell explosions.  Upper left panel:  $z =$ 7; upper right panel:  
$z =$ 10; lower left panel:  $z =$ 15; lower right panel:  $z =$ 20.  The dashed line at AB mag 32
indicates the \textit{JWST} NIRCam photometry limit.  The first peak at $z = $ 7 and 10 is due to 
the initial collision with the shell and cannot be seen at $z > $ 10.  The second peak is caused by 
breakout from the shell and is visible out to $z \gtrsim$ 20.}
\label{fig:NIR}
\end{figure*}

The first peak in the synthetic i-band light curves better matches the observations than the longer 
and dimmer late time peak that appears in both bands.  If these transients are Type IIn events it is 
likely that the initial collision with the shell is being observed, and breakout from the shell would be 
seen later if the surveys were extended.  We note that at $z = $ 2.2 the progenitors would be Pop 
II stars, not pristine Pop III stars, and the presence of even small amounts of dust or metals in the 
shell might have large effects on the opacity and further reduce the magnitude of the first peak in 
the i-band.  In addition, our simulation suite only explored the effect of changing the mass in simple 
analytic shells.  The light curves also depend on the energy of the SN explosion, the distance from 
the SN to the shell, and the thickness of the shell.  Therefore, it is not surprising that our 5 models
do not exactly match the observational data.  Future simulations are planned to explore a larger
parameter space for shell collision SNe in the local universe.

\section{NIR Light Curves}

We calculate NIR light curves for Pop III Type IIn SNe with the synthetic photometry code described 
in \citet{su11}.  We redshift each spectrum to the desired $z$ before removing the flux absorbed by 
intervening neutral hydrogen clouds with the method of \citet{madau95}.  Each spectrum is then 
dimmed by the required cosmological factors. Our algorithm linearly interpolates the least sampled 
data between the input spectrum and filter curve.  It has additional capabilities such as reddening by 
dust that are not used here.  

\begin{figure}
\plotone{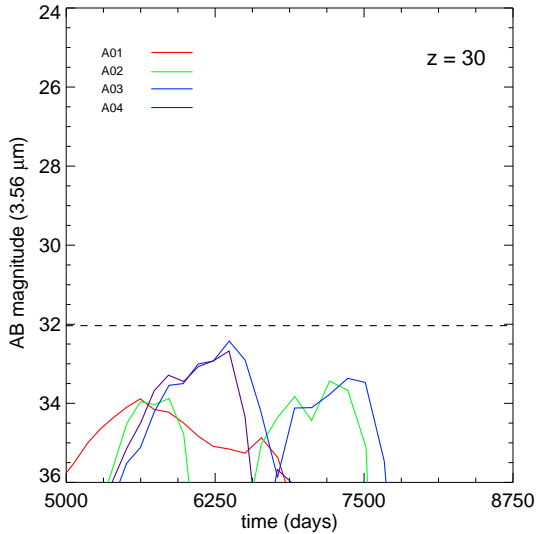}
\caption{NIR light curves for all 5 SNe at $z =$ 30.  The dashed line at AB mag 32 again denotes 
the \textit{JWST} NIRCam photometry limit.  } 
\label{fig:z30}
\end{figure}

We show NIR light curves for all 5 explosions at $z =$ 7, 10, 15, 20 and 30 in Figures \ref{fig:NIR}
and \ref{fig:z30}.  In each case we plot the NIR signal in the optimum filter for its detection, which in
all cases is redward of the Lyman limit at that redshift.  At $z =$ 7 and 10, two peaks are visible in 
the A01 - A04 explosions, a narrow, brighter peak at 500 - 1000 days and a dimmer, broader peak 
at 2000 - 2500 days.  The first is due to shock breakout from the outer surface of the shell.  The 
second occurs as the redshifted spectral peak evolves downward in wavelength through the NIR as 
the shell later expands and cools.  The first peak is present but not visible at higher redshifts.  Its 
luminosity does not change with shell mass, but more diffuse shells have broader peaks because 
photons from the shock can escape such shells before the shock, with lead times that are inversely 
proportional to the density of the shell.  The first peak appears at earlier times with low-mass shells 
because the shock breaks free of them sooner.  The A00 light curve does not exhibit this peak 
because there is no breakout from the shell. In contrast, the luminosity of the second peak rises with 
shell mass but its width is relatively uniform.  As expected, this second peak occurs at later times at 
higher redshifts.

With NIRCam photometry limits of AB magnitude 31 - 32, \textit{JWST} will be able to detect all five 
explosions out to $z \sim$ 20.  With proposed NIR detection limits of AB magnitude 26.5 - 27, 
\textit{WFIRST} and \textit{WISH} will only be able to observe such events out to $z \sim$ 7 - 10. At 
$z =$ 7 and 10 the first peak is visible for 250 - 500 days and the second peak is visible for $\sim$ 
1500 days.  At $z =$ 15 and 20 the second peak can be seen for $\sim$ 2000 days but is about a 
magnitude dimmer.  The peaks rise as quickly as they fall, with durations that are comparable to 
likely protogalactic survey times of 1 - 5 yr.  Because the spectrum of the shock becomes softer as 
it expands  and cools, its NIR light curve evolves on much shorter timescales than its redshifted 
bolometric luminosity.  These events will be easily recognizable as transients and distinguished 
from protogalaxies.  

\section{Conclusion}

Luminous Type IIn SNe will probe stellar populations at $z =$ 10 - 20, redshifts that complement 
those at which normal-luminosity CC and PI SNe can be detected ($z < $ 15 and $z \gtrsim$ 30, 
respectively).  They will not trace the first generation of stars, those that form at $z \sim$ 20 - 30, 
but they will be found during the rise of the Lyman-Werner background (10 $<z<$ 20) and in the 
first galaxies at $z \sim$ 10 - 15. The event rate of Type IIn SNe depends in part on what fraction 
of 20 - 40 \Ms\ Pop III stars die as compact blue giants or shed a common envelope in a binary, 
and may be low \citep{wa05,wl05,oet05,tfs07,wet08b,tss09,wet10,get10,maio11,hum12,jdk12,
wise12}.  For example, \citet{tet12} take this rate to be a few tenths of a percent of the total Type 
II SN rate.  However, such estimates may be too conservative for higher redshifts because the 
Pop III IMF is known to be top heavy and because the evolution of 20 - 40 \Ms\ primordial stars 
is still not fully understood.

The other challenge to observing such explosions is that they are too dim to be detected beyond
$z \gtrsim$ 7 in all-sky NIR surveys by \textit{WFIRST} or \textit{WISH}, whose wide fields of view
would otherwise compensate for the low SN IIn event rate.  This picture could change if the core
collapse event is more energetic than the 2.4 $\times$ 10$^{51}$ erg explosion considered here.
For example, \citet{moriya12} find that SN2006gy is best modeled by a 4 $\times$ 10$^{52}$ erg
hypernova explosion in a 15 \Ms\ circumstellar shell.  Such events, with or without shells, may be
visible in all-sky NIR campaigns at much higher redshifts and will be the focus of a future paper.

In our models we assume a uniform shell, but in reality it could be clumpy due to a variety of 
hydrodynamical instabilities.  If so, the ejecta would light up the shell unevenly, with brighter 
emission emanating from denser clumps than from diffuse regions.  It is not clear how the total
luminosity of the collision would change in such circumstances.  However, this scenario would 
occur less often in the primordial universe than today because there are no metals or dust to 
radiatively cool and fracture the shell.  Clumping could still occur if the medium into which the 
shell is ejected is not uniform or if the outflow itself is collimated. The former is less likely at high 
redshifts because the progenitor star usually drives all the gas from the halo in strong ionized 
flows, leaving behind diffuse uniform media in the vicinity of the star.  

Radio emission from these explosions may be visible at 21 cm by \textit{eVLA}, \textit{eMerlin} and 
the \textit{Square Kilometer Array} (\textit{SKA}). \citet{mw12} found that synchrotron emission from 
CC SNe at $z =$ 10 -- 20 will be detected by \textit{SKA} and that more energetic hypernovae at 
this epoch can be detected by existing facilities. Additional calculations are necessary to determine
if the collision of the ejecta with the shell enhances or quenches its radio emission.  Type IIn SNe
are not expected to imprint excess power on the CMB on small scales because unlike PI SNe, CC 
SNe are not sufficiently energetic to Comptonize large numbers of CMB photons \citep{oh03,wet08a}.

A small fraction of Pop III CC SNe may proceed as gamma ray bursts \citep[GRBs; e.g.,][]{bl06,
wang12}, driven either by the collapse of massive rapidly rotating stars \citep{suwa11,nsi12} or 
binary mergers between 20 - 50 \Ms\ stars \citep[e.g.,][]{fw98,fwh99,zf01,pasp07}.  This is 
reinforced by the fact that some Pop III stars have been found to form in binaries in numerical 
simulations \citep{turk09}.  While x-rays from these events could be detected by \textit{Swift} or 
next-generation missions such as the \textit{Joint Astrophysics Nascent Universe Satellite} 
\citep[\textit{JANUS}][]{mesz10,Roming08,burrows10}, it is more likely that their afterglows 
\citep{wet08c} will be detected in all-sky radio surveys by the \textit{eVLA}, \textit{eMERLIN} and 
\textit{SKA} \citep{ds11} due to their low event rates. If these events occur in dense circumstellar 
shells like those in our models, the shells may imprint distinct features on the afterglows 
\citep[e.g.,][]{met12a}.  We are now determining the afterglow signatures of Pop III GRBs in a 
variety of circumburst environments \citep{met12a,mes13a}.  

Pop III Type II SNe will completely outshine the primeval galaxies in which they occur because they
have comparatively few stars.  These events may reveal the existence of such galaxies when they 
might not otherwise have been detected by \textit{JWST} or future 30 m class telescopes such as 
the \textit{Giant Magellan Telescope} or the \textit{Thirty-Meter Telescope}.  Together with CC and 
PI SNe, Pop III shell-collision SNe will trace star formation rates and chemical enrichment in nascent 
galaxies.  These ancient explosions will soon open a new window on the high-redshift universe.

\acknowledgments

We thank the anonymous referee, whose comments improved the quality of this paper.  DJW is 
grateful for helpful discussions with Lucy Frey and Candace Joggerst and for support from the 
Bruce and Astrid McWilliams Center for Cosmology at Carnegie Mellon University.  MS thanks 
Marcia Rieke for making the NIRCam filter curves available and was partially supported by NASA 
JWST grant  NAG5-12458.  DEH acknowledges support from the National Science Foundation 
CAREER grant PHY-1151836.  Our work in part is based on observations obtained with MegaPrime
and MegaCam, a joint project of CFHT and CEA/IRFU, at the Canada-France-Hawaii Telescope 
(CFHT) which is operated by the National Research Council (NRC) of Canada, the Institut National 
des Science de l'Univers of the Centre National de la Recherche Scientifique (CNRS) of France, 
and the University of Hawaii.  Our work is also based in part on data products produced at Terapix 
available at the Canadian Astronomy Data Centre as part of the Canada-France-Hawaii Telescope 
Legacy Survey, a collaborative project of NRC and CNRS.  Work at LANL was performed under the 
auspices of the National Nuclear Security Administration of the U.S. Department of Energy at Los 
Alamos National Laboratory under Contract No. DE-AC52-06NA25396.  All RAGE and SPECTRUM 
calculations were performed on Institutional Computing (IC) and Yellow network platforms at LANL 
(Conejo, Lobo and Yellowrail).

\bibliographystyle{apj}
\bibliography{refs}

\end{document}